\documentclass[11pt]{article}
\pdfoutput=1

\usepackage{jheppub}
\bibstyle{JHEP}

\usepackage[english]{babel} 
\usepackage[utf8]{inputenc}  
\usepackage[T1]{fontenc}
\usepackage{lmodern}
\usepackage{amsmath}         
\usepackage{amsfonts}          
\usepackage{amssymb}
\usepackage{graphicx} 
\usepackage{bbold}
\usepackage[colorlinks=true,urlcolor=blue,linkcolor=blue,citecolor=blue,linktocpage=true]{hyperref}

\usepackage{tikz}
\usetikzlibrary{patterns,arrows,positioning,arrows.meta}
\tikzset{>={Stealth[width=1.5mm,length=1.5mm]}}

\newcommand{\figref}[1]	{{fig.~\ref{#1}}}
\newcommand{\secref}[1]	{{sec.~\ref{#1}}}
\newcommand{\cF}{{\mathcal{F}}}

\newcommand{\cL}{{\mathcal{L}}}
\newcommand{\cO}{{\mathcal{O}}}
\newcommand{\cR}{{\mathcal{R}}}

\newcommand{\CC}{{\mathbb{C}}}
\newcommand{\NN}{{\mathbb{N}}}
\newcommand{\ZZ}{{\mathbb{Z}}}
\def\({\left(}
\def\){\right)}

\newcommand{\Tr}{{\mathrm{Tr}}}
\newcommand{\fusionpolynomial}[4][rs]{{P^{#1}_{#2}\biggl[\arraycolsep=0pt\begin{array}{c}#3\\#4\end{array}\biggr]}}

\begin{document}

\title{Monodromy methods for torus conformal blocks and entanglement entropy at large central charge}

\author{Marius Gerbershagen}
\affiliation{Institut f{\"u}r Theoretische Physik und Astrophysik \\ and W\"urzburg-Dresden Cluster of Excellence ct.qmat, \\ Julius-Maximilians-Universit{\"a}t W{\"u}rzburg, Am Hubland, 97074 W\"urzburg, Germany}
\emailAdd{marius.gerbershagen@physik.uni-wuerzburg.de}

\abstract{
  We compute the entanglement entropy in a two dimensional conformal field theory at finite size and finite temperature in the large central charge limit via the replica trick.
  We first generalize the known monodromy method for the calculation of conformal blocks on the plane to the torus.
  Then, we derive a monodromy method for the zero-point conformal blocks of the replica partition function.
  We explain the differences between the two monodromy methods before applying them to the calculation of the entanglement entropy.
  We find that the contribution of the vacuum exchange dominates the entanglement entropy for a large class of CFTs, leading to universal results in agreement with holographic predictions from the RT formula.
  Moreover, we determine in which regime the replica partition function agrees with a correlation function of local twist operators on the torus.
}

\maketitle

\newpage

\section{Introduction}
\label{sec:introduction}

Entanglement entropy is a measure for the amount of entanglement between two parts of a quantum system.
It is defined as the von Neumann entropy of the reduced density matrix $\rho_A$ for a subsystem $A$.
In general, the entanglement entropy depends on details of the theory and state in question such as the spectrum and operator content.
However, certain universal features are common to all quantum field theories.
For example, the leading order divergence in the UV cutoff usually scales with the area of the boundary of the subregion $A$ \cite{Bombelli:1986rw,Srednicki:1993im}.
Conformal field theories in two dimensions admit more general universal features.
In particular, the entanglement entropy of a single interval $A$ at zero temperature is given by \cite{Calabrese:2004eu}
\begin{equation}
  S_A = \frac{c}{3} \log(l/\epsilon_\text{UV}),
\end{equation}
depending only on the central charge, irrespective of any other details such as the OPE coefficients or the spectrum of the theory.
For subsystems $A$ consisting of multiple intervals, the entanglement entropy is no longer universal for any CFT.
However, as shown in \cite{Hartman:2013mia}, in the semiclassical large central charge limit and at zero temperature, the entanglement entropy again becomes universal for a large class of conformal field theories.
These CFTs are characterized by a sparse spectrum of light operators and at most exponentially growing OPE coefficients.
By using conformal transformations, the universal results of \cite{Calabrese:2004eu,Hartman:2013mia} translate to the case of either finite temperature or finite size.
This publication is dedicated to the study of universal features of the entanglement entropy in a system with both finite size and finite temperature.

The computational approach most commonly used to determine entanglement entropies is the replica trick.
It is based on the calculation of the Rényi entropies
\begin{equation}
  S_A^{(n)} = \frac{1}{1-n} \log \Tr \rho_A^n,
\end{equation}
via a partition function $Z_n$ on a higher genus Riemann surface $\cR_n$ obtained by gluing $n$ copies of the complex plane cyclically together along the entangling interval $A$.
This partition function is then mapped to a correlation function of twist operators, i.e.~local operators with scaling dimension $h=\bar h=c/24(n-1/n)$ inserted at the endpoints of the entangling interval.
For a subsystem $A$ consisting of $N$ disjoint intervals, the correlation function contains $2N$ twist operator insertions.
Finally, the entanglement entropy is obtained by analytically continuing $n$ to the real numbers and taking the limit $n \to 1$.
The universality of the entanglement entropy for a single interval follows immediately from the universality of the two-point function in any conformal field theory \cite{Calabrese:2004eu}.
For multiple intervals, the Rényi entropy is mapped to a higher-point correlation function of twist operators which decomposes into a sum over conformal blocks.
The universality in the semiclassical limit observed in \cite{Hartman:2013mia} is explained by the fact that only a single conformal block (the vacuum block) contributes to the entanglement entropy.
More precisely, in the semiclassical limit of large central charge, the contribution of other conformal blocks is exponentially suppressed in the central charge, assuming the aforementioned restrictions on the theory, i.e.~a sparse spectrum of light operators and at most exponentially growing OPE coefficients.

In the case of a system with both finite size and finite temperature, the replica trick instructs us to calculate the partition function on a higher genus surface constructed by gluing $n$ copies of the torus along the entangling interval $A$.
In the small interval limit, this partition function can be obtained analogously to the zero temperature case as a correlation function on the torus of two (local) twist operators inserted at the endpoints of the entangling interval (see e.g.~\cite{Azeyanagi:2007bj,Calabrese:2009qy,Cardy:2014jwa,Chen:2016lbu}).
For large intervals, on the other hand, the replica partition function does not agree with a correlation function of local twist operators, as can be seen by the following argument.
It is well known that for a pure state $\rho$, the entanglement entropy for $A$ is equal to the entanglement entropy of the complement $A^c$, $S_A = S_{A^c}$.
However, for mixed states such as the thermal states described by the CFT on the torus, this property no longer holds.
Since the correlation function of local twist operators contains no information about whether we compute the entanglement entropy for $A$ or for $A^c$ (the location of the branch cuts between the twist operators is not fixed by the twist correlator), it cannot give the correct answer on the torus\footnote{In the case of a free CFT, a discrepancy between the correlator of local twist operators and the higher genus partition function has been observed in \cite{Lokhande:2015zma,Mukhi:2017rex} based on previous work \cite{Headrick:2012fk,Datta:2013hba,Chen:2015cna}.
In particular, it was found in \cite{Lokhande:2015zma} that the twist operator result is not modular covariant and violates Bose-Fermi equivalence.}.
This issue can be resolved for instance by defining non-local twist operators on the torus\footnote{We would like to thank the referee for pointing this out to us.} as in \cite{Chen:2014hta}, however we will not do this and instead phrase the calculation directly in terms of the replica partition function.
Therefore, in the following the term ``twist operator'' will always refer to a local operator of scaling dimension $h = \bar h = c/24(n-1/n)$.

A convenient way to calculate the entanglement entropy on the plane works by expanding the twist correlator in conformal blocks and calculating these using the well-known monodromy method first described in \cite{Zamolodchikov1987}.
This monodromy method is derived as follows.
By inserting a degenerate field into the correlation function, one obtains a differential equation for an auxiliary function $\Psi(z)$.
This differential equations contains derivatives of the sought after conformal block as accessory parameters.
The accessory parameters are fixed by demanding a certain monodromy of the solution $\Psi(z)$ around cycles which encircle a number of operator insertion points.
Which insertion points are encircled depends on the channel in which the correlation function is expanded.
We review this method in detail in \secref{sec:monodromy-method-plane} before generalizing it to the case of finite temperature and finite size.

A different perspective on this method was offered in \cite{Faulkner:2013yia}, where it was related to a uniformization problem on the replica surface $\cR_n$.
In general, a compact Riemann surface $\Sigma$ can be obtained as the quotient of the complex plane by a subgroup of $PSL(2,\CC)$ \cite{Zograf_1988}.
Thus, there exists a single valued map $w \to z$ from the complex plane to $\cR_n$.
This uniformization map is given as the quotient $w = \Psi_1(z)/\Psi_2(z)$ of two independent solutions of a differential equation.
It turns out that this differential equation is equal to the one from the monodromy method for the conformal blocks of the twist correlator on the plane \cite{Faulkner:2013yia}.
Using the equivalence between the CFT partition function and the gravitational action in the dual AdS space, this yields a proof of the RT formula at zero temperature \cite{Faulkner:2013yia}.
We use similar arguments from the uniformization problem on the replica surface of the torus to determine a monodromy method for the zero-point conformal blocks of the replica partition function at finite temperature.
This new monodromy method is closely connected to the monodromy method for the conformal blocks of the correlator of local twist operators on the torus, with the crucial difference being that the new method allows for choosing a larger set of cycles around which to impose the monodromy.
These new cycles are necessary to reproduce the $S_A \neq S_{A^c}$ property for thermal states.

In the context of the AdS/CFT correspondence, universal features of the entanglement entropy for holographic CFTs at large central charge are predicted by the Ryu-Takayanagi formula (RT formula for short) \cite{Ryu:2006bv}.
The RT formula states that the entanglement entropy of a subregion $A$ in the boundary field theory corresponds to the area of a minimal surface $\gamma_A$ in the bulk, anchored at $\partial A$ on the boundary of the AdS space.
Apart from reproducing the universal results obtained in \cite{Calabrese:2004eu,Hartman:2013mia}, the RT formula also predicts interesting universal features of the entanglement entropy in the case of both finite temperature and finite size.
In particular, there are two phase transitions \cite{Azeyanagi:2007bj}.
First, there is a Hawking-Page transition in the bulk from thermal AdS$_3$ to the BTZ black hole phase as the temperature increases.
This induces a corresponding phase transition in the entanglement entropy.
Second, the entanglement entropy in the BTZ phase also shows a phase transition as the size of the entangling interval increases.
We explain how these features appear from the CFT side.

Related work on the entanglement entropy in conformal field theories at finite size and finite temperature includes \cite{Chen:2014hta,Chen:2014ehg,Chen:2014unl,Barrella:2013wja,Chen:2015kua}.
\cite{Chen:2014hta,Chen:2014ehg,Chen:2014unl} is concerned with the entanglement entropy in various limits of high and low temperature or size of the entangling interval $A$, in which case universal results for arbitrary values of the central charge can be obtained.
In \cite{Barrella:2013wja,Chen:2015kua}, the holographic entanglement entropy for a single entangling interval on the boundary of a thermal AdS$_3$ space and the BTZ black hole was calculated using a monodromy method on the gravity side.
The monodromy method derived from the CFT side in this publication will turn out to be equivalent to the monodromy method on the gravity side used in \cite{Barrella:2013wja,Chen:2015kua}.
Related work on torus conformal blocks includes \cite{KashaniPoor:2012wb} which derived the monodromy method for the special case of one-point Virasoro conformal blocks on the torus and \cite{Alkalaev:2016fok,Alkalaev:2017bzx} which performed explicit calculations of one- and two-point Virasoro conformal blocks in various limits including the semiclassical one which we study in this publication.

Our paper is organized as follows.
In \secref{sec:monodromy-methods}, we derive the monodromy methods used in this publication.
After a review of the standard monodromy method for conformal blocks on the plane in \secref{sec:monodromy-method-plane}, we generalize to torus conformal blocks in \secref{sec:monodromy-method-torus}.
Sec.~\ref{sec:monodromy-method-replica-manifold} explains how to obtain a monodromy method for zero-point conformal blocks of the partition function on the replica surface and the differences between it and the monodromy method for conformal blocks on the torus.
Following this, we apply the newly derived monodromy methods to the calculation of the entanglement entropy in \secref{sec:EE-thermal-states}.
Assuming that the vacuum exchange dominates the partition function on the higher genus Riemann surface, we find universal results in agreement with the RT formula.
For the partition function on the replica surface, we find in particular agreement with the phase transition in the entanglement entropy at large interval size and high temperature.
This feature cannot be reproduced from the correlator of local twist operators.
We check the assumption on the dominance of the vacuum exchange numerically in \secref{sec:vacuum-block-dominance}.
Finally, we conclude with a brief discussion and outlook in \secref{sec:discussion}.

\section{Monodromy methods}
\label{sec:monodromy-methods}
This section contains an overview over the monodromy methods used in this publication.
We start with a review of the standard monodromy method for conformal blocks on the plane, then turn to the case of conformal blocks on the torus and finally explain how to derive a monodromy method for zero-point blocks of the partition function on the replica surface relevant to the computation of entanglement entropy on the torus.

\subsection{Conformal blocks on the plane}
\label{sec:monodromy-method-plane}
In this section, we review the monodromy method for the calculation of four-point semiclassical conformal blocks on the plane first derived in \cite{Zamolodchikov1986} (see also \cite{Harlow:2011ny} for a more detailed explanation).
The starting point of the derivation is the correlation function of four primary fields $\cO_i$
\begin{equation}
  \langle \cO_1(z_1,\bar z_1)\cO_2(z_2,\bar z_2)\cO_3(z_3,\bar z_3)\cO_4(z_4,\bar z_4) \rangle.
  \label{eq:4-point-function-plane}
\end{equation}
Let us parametrize the central charge as $c=1+6(b+1/b)^2$ and take the semiclassical limit $c \to \infty, b \to 0$ in which the conformal weights $h_i$ of the operators $\cO_i$ as well as the internal conformal weight $h_p$ scale proportional to the central charge.
In the correlation function \eqref{eq:4-point-function-plane}, we insert a degenerate operator $\Psi(z,\bar z)$ with conformal weight $h_\Psi = -1/2-3b^2/4 \sim \cO(c^0)$ obeying
\begin{equation}
  \left(\hat L_{-2} + \frac{1}{b^2}\hat L_{-1}^2\right)\Psi(z,\bar z) = 0
\end{equation}
From the conformal Ward identities, this yields the following differential equation known as the decoupling equation,
\begin{equation}
  \left[
    \frac{1}{b^2}\partial_z^2 + \sum_i \left(\frac{h_i}{(z-z_i)^2} + \frac{\partial_{z_i}}{z-z_i}\right)
  \right] \langle \cO_1 \cO_2 \Psi \cO_3 \cO_4 \rangle = 0.
  \label{eq:decoupling-eq-plane-1}
\end{equation}
To get to the $s$-channel conformal block, we then insert the operator product expansion
\begin{equation}
  \cO_1(z_1,\bar z_1)\cO_2(z_2,\bar z_2) = \sum_p C^p_{21} \sum_{k,\bar k} (z_2-z_1)^{h_p-h_1-h_2+|k|} (\bar z_2-\bar z_1)^{\bar h_p-\bar h_1-\bar h_2+|\bar k|} \beta^{pk}_{21}\beta^{p\bar k}_{21} \cO_p^{\{k,\bar k\}}(z_1,\bar z_1)
\end{equation}
into the correlation function which yields terms containing $\langle \cO_p^{\{k,\bar k\}} \Psi \cO_3 \cO_4 \rangle$.
At large central charge, these terms can be approximated by
\begin{equation}
  \langle \cO_p^{\{k,\bar k\}} \Psi \cO_3 \cO_4 \rangle \approx \Psi_p \langle \cO_p^{\{k,\bar k\}} \cO_3 \cO_4 \rangle,
  \label{eq:psi_p-approximation}
\end{equation}
where $\Psi_p$ is defined by
\begin{equation}
  \Psi_p = \frac{\langle \cO_p \Psi \cO_3 \cO_4 \rangle}{\langle \cO_p \cO_3 \cO_4 \rangle}.
\end{equation}
This can be shown by employing the form of $\langle \cO_p^{\{k,\bar k\}} \Psi \cO_3 \cO_4 \rangle$ in terms of a string of differential operators $\cL_{k_i},\bar \cL_{k_i}$ acting on $\langle \cO_p \Psi \cO_3 \cO_4 \rangle$, where
\begin{equation}
  \cL_{-k_i}^\Psi = -\sum_{j=3,4,\Psi} \left( \frac{(1-k_i)h_j}{(z_j-z_1)^{k_i}} + \frac{1}{(z_j-z_1)^{k_i-1}} \partial_{z_j}\right).
\end{equation}
Now $\langle \cO_p \Psi \cO_3 \cO_4 \rangle$ scales as $e^{-c/6 S_\text{cl.}}$ in the semiclassical limit where $S_\text{cl.} \sim \cO(c^0)$ while $\Psi_p \sim \cO(c^0)$ and $h_\Psi \sim \cO(c^0)$.
Hence, we can neglect the derivatives acting on $\Psi_p$ and on the $h_\Psi$ term to obtain \eqref{eq:psi_p-approximation} in the leading order in $c$.
Now, use a conformal transformation to send $z_1 \to 0$, $z_3 \to 1$, $z_4 \to \infty$ and $z_2$ to the cross ratio $x$.
This implies
\begin{equation}
  \langle \cO_1 \cO_2 \Psi \cO_3 \cO_4 \rangle \approx \sum_p \Psi_p(z,x,\bar x) C^p_{21} C^p_{43} \cF^p_{12,34}(x) \bar \cF^p_{12,34}(\bar x),
\end{equation}
where $\cF^p_{12,34}(x)$ is the desired conformal block which in the semiclassical limit scales as $\cF^p_{12,34}(x) \sim e^{-\frac{c}{6} f_\text{cl.}(x)}$ as was conjectured in \cite{Zamolodchikov1986} and recently shown in \cite{Besken:2019jyw}.
The semiclassical conformal block $f_\text{cl.}$ depends only on the cross ratio $x$ and on $b^2 h_i,b^2 h_p$.
The decoupling equation \eqref{eq:decoupling-eq-plane-1} then yields at leading order in $c$
\begin{equation}
  \left[
    \partial_z^2 + \sum_i \left(\frac{b^2 h_i}{(z-z_i)^2} - \frac{\partial_{z_i} f_\text{cl.}(x)}{z-z_i}\right)
  \right] \Psi_p = 0.
\end{equation}
There is one separate decoupling equation for each term in the sum over $p$ since generically, each term has a different monodromy and thus must vanish separately.
All terms involving derivatives of $\Psi_{p}$ vanish to leading order due to $\Psi_{p} \sim \cO(c^0)$.
From the expression for the cross ratio $x = \frac{(z_1-z_2)(z_4-z_3)}{(z_4-z_2)(z_1-z_3)}$, we obtain linear relations among the $\partial_{z_i} f_\text{cl.}$
\begin{equation}
  \sum_i \partial_{z_i} f_\text{cl.} = \sum_i (z_i \partial_{z_i} f_\text{cl.} - b^2 h_i) = \sum_i (z_i^2 \partial_{z_i} f_\text{cl.} - 2 z_i b^2 h_i) = 0.
\end{equation}
These follow from $\partial_{z_i} f_\text{cl.} = (\frac{\partial x}{\partial z_i}) \partial_x f_\text{cl.}$ and $\sum_i \frac{\partial x}{\partial z_i} = \sum_i \frac{\partial x}{\partial z_i} z_i = \sum_i \frac{\partial x}{\partial z_i} z_i^2 = 0$ as can easily be shown from the definition of $x$ and the conformal transformation properties of correlation functions of primary operators.
This yields the final form of the decoupling equation,
\begin{equation}
  \left[
    \partial_z^2 + \frac{b^2 h_1}{z^2} + \frac{b^2 h_2}{(z-x)^2} + \frac{b^2 h_3}{(z-1)^2} - \frac{b^2(h_1+h_2+h_3-h_4)}{z(z-1)} + \frac{x(1-x)\partial_x f_\text{cl.}}{z(z-x)(z-1)}
  \right] \Psi_p = 0.
  \label{eq:decoupling-eq-plane-3}
\end{equation}
To obtain $f_\text{cl.}$ from this equation, we use the fact that the solutions $\Psi_p$ must have a certain monodromy when $z$ is taken in a loop around $0,x$.
This monodromy can be derived from the decoupling equation of $\langle \cO_p \Psi \cO_3 \cO_4 \rangle$,
\begin{equation}
  \biggl[
  \frac{1}{b^2} \partial_z^2 + \left(\frac{h_p}{(z-z_1)^2} + \frac{1}{z-z_1}\partial_{z_1}\right) + \sum_{i=3,4} \left(\frac{h_i}{(z-z_i)^2} + \frac{1}{z-z_i}\partial_{z_i}\right)
  \biggr] \langle \cO_p \Psi \cO_3 \cO_4 \rangle = 0.
  \label{eq:decoupling-equation-correlator-Psi-O_p}
\end{equation}
As $z \to z_1$, the leading coefficient of the OPE between $\Psi$ and $\cO_p$ is given by $(z-z_1)^\kappa \cO_p'(z_1)$ where $\kappa$ can be determined by inserting this coefficient into \eqref{eq:decoupling-equation-correlator-Psi-O_p},
\begin{equation}
  \begin{aligned}
    \biggl[
    &\frac{1}{b^2}\kappa(\kappa-1) (z-z_1)^{\kappa-2} + \sum_{i=3,4} \left( \frac{h_i}{(z-z_i)^2} + \frac{1}{z-z_i}\partial_{z_i}\right)(z-z_1)^\kappa\\
    &+ h_p(z-z_1)^{\kappa-2} - \kappa (z-z_1)^{\kappa-2}
    \biggr] \langle \cO_p' \cO_3 \cO_4 \rangle = 0.
  \end{aligned}
\end{equation}
The leading contribution in $z \to z_1$ is given by
\begin{equation}
  \biggl[ \frac{1}{b^2}\kappa(\kappa-1) + h_p - \kappa \biggl] (z-z_1)^{\kappa-2} = 0.
\end{equation}
Thus as $b^2 \to 0$ and $z \to z_1$,
\begin{equation}
  \kappa = \frac{1}{2} \left(1 \pm \sqrt{1-4h_pb^2}\right) \text{~~and~~} \Psi_{p} \sim (z-z_1)^{\frac{1}{2}(1 \pm \sqrt{1-4h_pb^2})}.
\end{equation}
Therefore, the monodromy matrix around $0,x$ is given by
\begin{equation}
  M_{0,x} =
  \left(
    \begin{array}{cc}
      e^{i\pi(1 + \sqrt{1-4h_pb^2})} & 0\\
      0 & e^{i\pi(1 - \sqrt{1-4h_pb^2})}\\
    \end{array}
  \right).
\end{equation}
The trace of the monodromy matrix, which is independent of the basis in which the two solutions of \eqref{eq:decoupling-eq-plane-3} are decomposed, is given by
\begin{equation}
  \Tr M_{0,x} = -2\cos\(\pi\sqrt{1-4h_pb^2}\).
  \label{eq:monodromy-Psi_p}
\end{equation}
Thus, the torus conformal block can be extracted from \eqref{eq:decoupling-eq-plane-3} by choosing $\partial_x f_\text{cl.}$ such that the monodromy of the solution $\Psi_{p}$ around a loop enclosing $z_1$ and $z_2$ is given by \eqref{eq:monodromy-Psi_p} \footnote{The loop needs to enclose both $z_1$ and $z_2$ in order for the OPE between $\cO_1(z_1)$ and $\cO_2(z_2)$ to converge.}.
Finally, the conformal block is obtained by integrating the chosen $\partial_x f_\text{cl.}$.

The four point conformal block in other channels is obtained from the same decoupling equation by imposing different monodromy conditions.
For example, for the $t$-channel block we impose the monodromy condition around the insertion points of $\cO_2$ and $\cO_3$, $\Tr M_{1,x} = -2\cos\(\pi\sqrt{1-4h_pb^2}\)$.
Higher point conformal blocks on the plane are computed from similar monodromy methods derived analogously to the four-point case.
For $n$ point blocks, the decoupling contains $n-3$ independent derivatives fixed by $n-3$ monodromy conditions around the operator insertion points which are contracted in the OPE.

\subsection{Conformal blocks on the torus}
\label{sec:monodromy-method-torus}
We now continue with the derivation of a monodromy method for conformal blocks on the torus.
The derivation of this monodromy method proceeds in a very similar way to the one on the plane.
We illustrate the derivation using the two-point function on the torus
\begin{equation}
  \langle \cO_1(z_1) \cO_2(z_2) \rangle_\tau = \Tr[e^{2\pi i\tau(L_0 - c/24)}e^{-2\pi i\bar\tau(\bar L_0 - c/24)} \cO_1(z_1) \cO_2(z_2)],
  \label{eq:2-point-function-torus}
\end{equation}
however conformal blocks for other correlation functions on the torus are obtained in a similar fashion, as we briefly discuss as the end of this section.
The modular parameter of the torus is denoted by $\tau$, related to the inverse temperature by $\beta = -2\pi i\tau$.
We also introduce the parameter
\begin{equation}
  Q = e^{-\beta} = e^{2\pi i \tau},
\end{equation}
written with an uppercase $Q$ instead of the standard lowercase $q$ to distinguish it from the conformal dimension $h_q$ of the internal index of the conformal block which we are about to derive.

As on the plane, we insert the degenerate operator $\Psi(z,\bar z)$ into \eqref{eq:2-point-function-torus}.
To derive the corresponding decoupling equation, we use the conformal Ward identity on the torus \cite{Eguchi:1986sb}~\footnote{Note that \cite{Eguchi:1986sb} uses a convention where correlation functions on the torus are normalized by the inverse of the partition function $Z(\tau)$ and thus the expression for the conformal Ward identity there contains an additional term $(2\pi i\partial_\tau Z(\tau)) \langle \prod_i \cO_i(z_i) \rangle_\tau$.},
\begin{equation}
  \begin{aligned}
    \langle T(z) \prod_i \cO_i(z_i) \rangle_\tau = \left[\sum_i\bigl(h_i \left(\wp(z-z_i) + 2\eta_1\right) + \left(\zeta(z-z_i) + 2\eta_1z_i\right)\partial_{z_i}\bigr) + 2\pi i\partial_\tau\right] \langle \prod_i \cO_i(z_i) \rangle_\tau.
  \end{aligned}
  \label{eq:Ward-identity-torus}
\end{equation}
Here, $z \sim z + 1 \sim z + \tau$ are the coordinates on the torus with modular parameter $\tau$ and $\wp(z),\zeta(z)$ denote Weierstraß elliptic functions with associated $\eta_1$ parameter (see App.~\ref{app:conventions} for more details on the Weierstraß functions).
Using the definition of the Virasoro generators,
\begin{equation}
  \hat L_{-n} \Psi(z) = \int \frac{dw}{2\pi i} \frac{1}{(w-z)^{n-1}} T(w) \Psi(z).
\end{equation}
and the conformal Ward identity \eqref{eq:Ward-identity-torus}, we see that
\begin{equation}
  \begin{aligned}
    \langle \prod_i \cO_i(z_i) \hat L_{-2}\Psi(z) \rangle_\tau = 
    \biggl[
    &\sum_i \left(h_i(\wp(z-z_i) + 2\eta_1) + (\zeta(z-z_i) + 2\eta_1 z_i)\partial_{z_i}\right)\\
    &+ 2 \eta_1 z\partial_z
    + 2 h_\Psi \eta_1
    + 2\pi i\partial_\tau
    \biggr]
    \langle \prod_i \cO_i(z_i) \Psi(z) \rangle_\tau
  \end{aligned}
\end{equation}
and
\begin{equation}
  \langle \prod_i \cO_i(z_i) \hat L_{-1}\Psi(z) \rangle_\tau = \partial_z \langle \prod_i \cO_i(z_i) \Psi(z) \rangle_\tau.
\end{equation}
Thus $\langle \prod_i \cO_i(z_i) \Psi(z) \rangle_\tau$ obeys the decoupling equation
\begin{equation}
  \begin{aligned}
    \biggl[
    &\frac{1}{b^2} \partial_z^2 + \sum_i \left(h_i(\wp(z-z_i) + 2\eta_1) + (\zeta(z-z_i) + 2\eta_1 z_i)\partial_{z_i}\right)\\
    &+ 2 \eta_1 z\partial_z
    + 2 h_\Psi \eta_1
    + 2\pi i\partial_\tau
    \biggr] \langle \prod_i \cO_i(z_i) \Psi(z) \rangle_\tau = 0.
  \end{aligned}
  \label{eq:decoupling-equation-torus-1}
\end{equation}
To relate $\langle \cO_1(z_1) \cO_2(z_2) \Psi(z) \rangle_\tau$ to a conformal block, we decompose the trace over states into contributions from a primary $\cO_q$ and its descendants and insert the appropriate OPE contractions.
\begin{figure}
  \centering
  \begin{tikzpicture}[scale=1.1]
    \begin{scope}[shift={(-4,0)}]
      \draw (-1,1) node[left] {{1}} -- (0,0);
      \draw (-1,-1) node[left] {{2}} -- (0,0);
      \draw (0,0) -- node[midway,below] {{$p$}} (1.5,0);
      \draw (2.5,0) circle[radius=1cm];
      \draw (3.5,0) node[right] {{$q$}};
    \end{scope}
    \begin{scope}[shift={(3,0)}]
      \draw (-1,1) node[left] {{1}} -- (-0.75*0.71,0.75*0.71);
      \draw (-1,-1) node[left] {{2}} -- (-0.75*0.71,-0.75*0.71);
      \draw (-0.75,0) node[left] {{$p$}};
      \draw (0,0) circle[radius=0.75cm];
      \draw (0.75,0) node[right] {{$q$}};
    \end{scope}
  \end{tikzpicture}
  \caption{Conformal blocks for the two point function on the torus. Left: OPE channel, right: projection channel.}
  \label{fig:two-point-torus-blocks}
\end{figure}
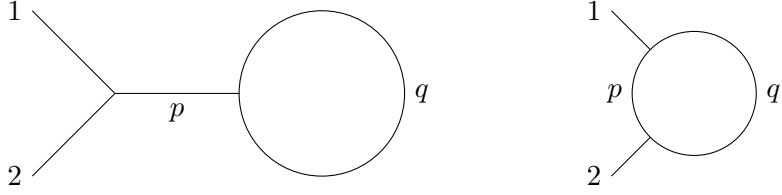
For the two-point function, there are two possible channels (see \figref{fig:two-point-torus-blocks}).
The \textit{projection block} is obtained by OPE contracting $\cO_2$ and $\cO_q$.
On the other hand, for the \textit{OPE block} we contract $\cO_2$ and $\cO_1$,
\begin{equation}
  \begin{aligned}
    &\langle \cO_1(z_1) \cO_2(z_2) \Psi(z) \rangle_\tau = \sum_q \sum_{l} Q^{h_q - c/24 + |l|} \langle \cO_q^{\{l\}}(z_0) \cO_1(z_1) \cO_2(z_2) \Psi(z) \cO^{\{l\}}_q(z_\infty) \rangle \text{(c.c.)}\\
    &= \sum_{p,q} C^p_{21} \sum_{k,l} (z_2 - z_1)^{h_p-h_1-h_2 + |k|} Q^{h_q - c/24 + |l|} \beta^{pk}_{21} \langle \cO^{\{l\}}_q(z_0) \cO^{\{k\}}_p(z_1) \Psi(z) \cO^{\{l\}}_q(z_\infty) \rangle \text{(c.c.)},
  \end{aligned}
\end{equation}
where (c.c.) denotes schematically the antiholomorphic parts of the expression and $z_0 \to -i\infty$ while $z_\infty \to +i\infty$.
We define
\begin{equation}
  \Psi_{pq} = \frac{\langle \cO_q(z_0) \Psi(z) \cO_p(z_1) \cO_q(z_\infty) \rangle}{\langle \cO_q(z_0) \cO_p(z_1) \cO_q(z_\infty) \rangle}.
\end{equation}
As on the plane, in the large $c$ limit
\begin{equation}
  \langle \cO^{\{l\}}_q(z_0) \cO^{\{k\}}_p(z_1) \Psi(z) \cO^{\{l\}}_q(z_\infty) \rangle \text{(c.c.)} \approx \Psi_{pq} \langle \cO^{\{l\}}_q(z_0) \cO^{\{k\}}_p(z_1) \cO^{\{l\}}_q(z_\infty) \rangle \text{(c.c.)}
\end{equation}
This yields
\begin{equation}
  \langle \cO_1 \cO_2 \Psi \rangle_\tau \approx \sum_{p,q} C^p_{21} C^q_{pq} \Psi_{pq} \cF_{21,pq} \bar\cF_{21,pq},
\end{equation}
where $\cF_{21,pq}$ is the conformal block which we want to compute.
Assuming that exponentiation of the conformal blocks in the semiclassical limit holds, $\cF_{21,pq} \sim e^{-c/6\,f_\text{cl.}}$, and using that $\partial_{z_1} f_\text{cl.} = - \partial_{z_2} f_\text{cl.}$, we obtain
\begin{equation}
  \biggl[
  \partial_z^2 + \sum_{i=1,2} 
  \left(
    b^2h_i(\wp(z-z_i) + 2\eta_1)
    + \partial_{z_2} f_\text{cl.} (-1)^{i+1}(\zeta(z-z_i) + 2\eta_1 z_i)
  \right)
  - 2\pi i\partial_\tau f_\text{cl.}
  \biggr]\Psi_{pq} = 0.
  \label{eq:decoupling-equation-torus-2}
\end{equation}
From the definition of $\Psi_{pq}$ we derive the monodromy conditions in the same way as on the plane.
For the OPE block, these are
\begin{equation}
  \Tr M_{z_1,z_2} = -2\cos(\pi\sqrt{1-4h_pb^2}) ,~ \Tr M_{z_0} = -2\cos(\pi\sqrt{1-4h_qb^2}).
\end{equation}
The subscripts of the monodromy matrices show around which cycles the monodromy is taken.
The derivation for the projection block works analogously.
Here we contract $\cO_q(z_0) \cO_2(z_2)$:
\begin{equation}
  \begin{aligned}
    &\langle \cO_1(z_1) \cO_2(z_2) \Psi(z) \rangle_\tau\\
    &= \sum_{p,q} C^p_{2q} \sum_{k,l} (z_2-z_0)^{h_p-h_q-h_2 + |k|} Q^{h_q - c/24 + |l|} \beta^{pk}_{2q} \langle \cO^{\{k\}}_p(z_0) \cO_1(z_1) \Psi(z) \cO^{\{l\}}_q(z_\infty) \rangle \text{(c.c.)}.
  \end{aligned}
\end{equation}
Using $\Psi_{pq}$ defined by
\begin{equation}
  \Psi_{pq} = \frac{\langle \cO_p(z_0) \Psi(z) \cO_1(z_1) \cO_q(z_\infty) \rangle}{\langle \cO_p(z_0) \cO_1(z_1) \cO_q(z_\infty) \rangle}
\end{equation}
and related to the two point correlator by
\begin{equation}
  \langle \cO_1 \cO_2 \Psi \rangle_\tau \approx \sum_{p,q} C^p_{2q} C^q_{1p} \Psi_{pq} \cF_{2q,1p} \bar\cF_{2q,1p}
\end{equation}
we find the same decoupling equation \eqref{eq:decoupling-equation-torus-2}.
However, the monodromy conditions differ.
They are given by
\begin{equation}
  \Tr M_{z_0,z_2} = -2\cos(\pi\sqrt{1-4h_pb^2}) ,~ \Tr M_{z_\infty} = -2\cos(\pi\sqrt{1-4h_qb^2}).
\end{equation}

To solve the decoupling equation, it is useful to perform a coordinate transformation $u = e^{-2\pi i z}$.
Using the transformation of primary operators under conformal transformations as well as the series representations of the Weierstrass elliptic functions from app.~\ref{app:conventions}, the decoupling equation becomes
\begin{equation}
  \begin{aligned}
    \biggl[&\partial_u^2 + y (b^2h_2 - (1+y)\partial_y f_\text{cl.})\sum_{m=-\infty}^\infty \frac{Q^m}{u(u - Q^m)(u - Q^m (1 + y))} + \frac{1/4 - Q\partial_Q f_\text{cl.}}{u^2}\\
    &+ b^2h_1\sum_{m=-\infty}^\infty \frac{Q^m}{u(u - Q^m)^2} + b^2h_2\sum_{m=-\infty}^\infty \frac{Q^m(1+y)}{u(u - Q^m(1+y))^2}\biggr] \Psi_{pq} = 0,
  \end{aligned}
  \label{eq:decoupling-equation-torus-3}
\end{equation}
where we have chosen w.l.o.g.~$z_1=0$ and $e^{-2\pi iz_2} = 1+y$.
In these coordinates, the monodromy conditions become
\begin{equation}
  \begin{aligned}
    \Tr M_{1,1+y} &= -2\cos(\pi\sqrt{1-4h_pb^2}) &,~ \Tr M_{0} =& -2\cos(\pi\sqrt{1-4h_qb^2}) &\text{(OPE block)}\\
    \Tr M_{0,1+y} &= -2\cos(\pi\sqrt{1-4h_pb^2}) &,~ \Tr M_{\infty} =& -2\cos(\pi\sqrt{1-4h_qb^2}) &\text{(projection block)}\\
  \end{aligned}
\end{equation}
This representation of the decoupling equation is immediately applicable for the calculation of the OPE block, which is defined through a series expansion in $y$ and $Q$.
Using this series expansion as well as a WKB approximation for large $h_p,h_q$, \eqref{eq:decoupling-equation-torus-3} can be solved order by order.
For example, to first order in $y$ and $Q$ we get
\begin{equation}
  f_\text{cl.}^\text{OPE} = -b^2 (h_p-h_1-h_2) \log y - (b^2 h_q - 1/4) \log Q + \frac{1}{2} b^2(h_p + h_2 - h_1) y - b^2\frac{h_p^2}{2 h_q} Q + ...
\end{equation}
The projection block, on the other hand, can be expanded in a series in $q_1 = Q/(1+y)$ and $q_2 = 1+y$.
The decoupling equation can then be solved in the same way as for the OPE block order by order in $q_1$ and $q_2$.
For example, to first order in $q_1$ and $q_2$ we obtain
\begin{equation}
  \begin{aligned}
    f_\text{cl.}^\text{projection} = &-(b^2(h_p - h_2) - 1/4) \log q_2 - (b^2 h_q - 1/4) \log q_1\\
    &- b^2\frac{(h_1-h_p+h_q)(h_2-h_p+h_q)}{2h_q}q_1 - b^2\frac{(h_1+h_p-h_q)(h_2+h_p-h_q)}{2h_p}q_2 + ...
  \end{aligned}
\end{equation}
We have checked that the results for both the OPE and the projection block are in agreement with the recursion formulas derived in \cite{Cho:2017oxl} (see app.~\ref{app:recursion-relations} for detailed expressions) as well as explicit calculations up to third order.

\medskip

It is clear that the above derivation can be easily generalized to other conformal blocks on the torus.
The simplest case is the zero-point block on the torus, i.e.~the Virasoro character.
Performing a similar derivation as above or equivalently taking the limit $h_{1,2,p} \to 0$ in \eqref{eq:decoupling-equation-torus-3}, we arrive at the following decoupling equation
\begin{equation}
  \left[\partial_u^2 + \frac{1/4 - Q \partial_Q f_\text{cl.}}{u^2}\right] \Psi_q = 0,
\end{equation}
together with the monodromy condition $\Tr M_0 = -2\cos(\pi\sqrt{1-4h_qb^2})$.
In this case, we can give the full solution.
The decoupling equation is solved by $\Psi_q = u^{1/2 \pm \sqrt{Q \partial_Q f_\text{cl.}}}$, from which we obtain $f_\text{cl.} = (1/4 - b^2 h_q)\log Q$ which correctly reproduces the leading order contribution in $c$ of the Virasoro character $\chi_q = \frac{1}{\eta(\tau)} Q^{h_q - (c - 1)/24} \sim e^{-c/6 f_\text{cl.}}$.
For a general $n$-point conformal block, the decoupling equation is given by
\begin{equation}
  \biggl[
  \partial_z^2 + \sum_{i=1}^n
  \left(
    b^2h_i(\wp(z-z_i) + 2\eta_1)
    + \partial_{z_i} f_\text{cl.}(\zeta(z-z_i) + 2\eta_1 z_i)
  \right)
  - 2\pi i\partial_\tau f_\text{cl.}
  \biggr]\Psi = 0,
  \label{eq:decoupling-equation-torus-general}
\end{equation}
and there are $n$ monodromy conditions around non-trivial cycles determined by the OPE contractions.
By conformal transformations, the insertion point of one of the operators can be fixed, for example to $z_1 \to 0$.
Then, there are $n$ independent accessory parameters $\partial_\tau f_\text{cl.}$ and $\partial_{z_i} f_\text{cl.}$ for $i \geq 2$ fixed by these monodromy conditions.

\subsection{Partition function on the replica surface}
\label{sec:monodromy-method-replica-manifold}
We now turn to the computation of the partition function $Z_n$ on the replica surface $\cR_n$.
In general, the partition function on any higher genus Riemann surface can be expanded in zero-point conformal blocks, which can again be calculated via a monodromy method.
This monodromy method can be derived by inserting the degenerate operator directly on the higher genus Riemann surface -- in contrast to the last section, where we inserted it in a correlation function on the torus -- and inserting projection operators in the appropriate places.
The difficulty of this approach is of course that deriving the decoupling equation for an arbitrary Riemann surface is quite hard.
However, we will see that things simplify for the special higher genus surface that we are interested in, that is the replica surface $\cR_n$.
Assuming that the dominant contribution to the partition function depends only on the temperature and size of the entangling interval and not on any other moduli of $\cR_n$, we find the same decoupling equation as for the twist operator correlator on the torus.
The difference to the last section lies in the monodromy conditions.
The zero-point block on $\cR_n$ admits more general monodromy conditions (corresponding to different channels) than the conformal block on the torus.
One of these more general monodromy conditions will give the dominant contribution to the entanglement entropy for large intervals.

Before deriving the decoupling equation on $\cR_n$, we collect some facts about the topology and moduli of $\cR_n$.
For simplicity, we specialize again to the single interval case.
The replica surface is given by $n$ copies of a torus with modular parameter $\tau$, joined at a branch cut along the entangling interval $A$.
We use coordinates $z,\bar z$ to parametrize $\cR_n$ with identifications $z \sim z + 1$ and $z \sim z + \tau$.
In these coordinates, $\cR_n$ is described by a branched cover of the torus with branch points located at $z = z_{1,2} + k + l\tau$ for $k,l \in \ZZ$.
Near the branch points, the covering map is given by $y^n \propto (z-z_1-k-l\tau)$ and $y^n \propto 1/(z-z_2-k-l\tau)$.
The genus of $\cR_n$ is then obtained by the Riemann-Hurwitz theorem.
The ramification index at each branch point is equal to $n$, yielding $g = n$.
Since the Euler characteristic is $\chi < 0$, there are no conformal Killing vectors.
This implies by the Riemann-Roch theorem that there exist $3(n-1)$ holomorphic quadratic differentials $\omega_{zz}^{(i)}$ parametrizing deformations of the complex structure of the Riemann surface.
The $\omega_{zz}^{(i)}$ are meromorphic doubly periodic functions that are regular everywhere on the covering surface, i.e.~$\omega_{yy}^{(i)} dy^2 = \omega_{zz}^{(i)}\bigl(\frac{dz}{dy}\bigr)^2 dz^2$ is non-singular for all $y$.
Simple examples include $\omega_{zz}^{(1)} = \text{const.}$ which is trivially regular and doubly periodic as well as $\omega_{zz}^{(2)} = \zeta(z-z_1) - \zeta(z-z_2) + 2\eta_1(z_1-z_2)$.
$\omega_{zz}^{(2)}$ is regular since near $z=z_1 + k + l\tau$ we have $\omega_{yy}^{(2)} \propto y^{n-2}$ which is regular at $y=0$ for $n \geq 2$.
Near $z=z_2 + k + l\tau$, regularity can be shown in an analogous way.
In fact, these two examples are the only ones relevant for the following arguments since they are the only ones that respect the $\ZZ_n$ replica symmetry permuting the different copies of the torus with each other\footnote{This can be seen as follows. The replica symmetry acts as $y \to y e^{2\pi i/n}$. Therefore, only the $\omega_{zz} \sim (z-z_i)^{\alpha_i}$ with $\alpha_i \in \ZZ$, $i=1,2$ are invariant under this symmetry. The case $\alpha_i < -1$ is singular at $z=z_i$. $\alpha_i > 0$ is singular at some other point since any non-constant elliptic function has at least two poles inside the fundamental parallelogram, which lead to singularities in $\omega_{yy}$. This leaves only $\alpha_i = 0,-1$ which are the two examples described above.}.

The derivation of the decoupling equation on the replica surface then proceeds in a similar fashion as in the previous section.
Assuming exponentiation of the zero-point block in the semiclassical limit, the conformal Ward identities for a general Riemann surface \cite{Eguchi:1986sb} imply a decoupling equation of the form
\begin{equation}
  \left[\partial_z^2 + \langle T_{zz} \rangle + \sum_{i=1}^{3(n-1)} \omega_{zz}^{(i)} \partial_{w_i} f_\text{cl.}\right] \Psi(z) = 0,
  \label{eq:decoupling-eq-replica-manifold}
\end{equation}
where $w_i$ are the modular parameters associated to $\omega_{zz}^{(i)}$.
$\langle T_{zz} \rangle$ is the expectation value of the energy momentum tensor.
It can be derived along the lines of \cite{Faulkner:2013yia}: $\langle T_{zz} \rangle$ transforms with a Schwarzian derivative,
\begin{equation}
 \langle T_{yy} \rangle = \left(\frac{\partial z}{\partial y}\right)^2 \langle T_{zz} \rangle + \frac{nc}{12} \{z,y\},
\end{equation}
and $\langle T_{yy} \rangle$ must be regular.
The Schwarzian derivative term comes with a $nc/12$ prefactor since the stress-energy tensor on the replica surface is given as the sum of the stress-energy tensors of the $n$ tori.
Therefore, the Schwarzian for the transformation of the stress-energy tensor on the replica surface is given by the sum of $n$ identical Schwarzian terms with prefactor $c/12$.
Together with the requirement that $\langle T_{zz} \rangle$ be doubly periodic, regularity of $\langle T_{yy} \rangle$ implies
\begin{equation}
  \langle T_{zz} \rangle = \frac{c}{24} \left(n-\frac{1}{n}\right) \sum_i (\wp(z-z_i) + 2\eta_1).
\end{equation}
The $1/(z-z_i)^2$ poles in $\wp(z-z_i)$ give a $1/y^2$ contribution to $\langle T_{yy} \rangle$ that cancels with the Schwarzian derivative term\footnote{For ease of comparison with the previous section we have also added a constant term $\frac{c}{24} \left(n-\frac{1}{n}\right) 4\eta_1$ to $\langle T_{zz} \rangle$ which is not strictly necessary for regularity and could be absorbed into the prefactor of $\omega_{zz}^{(1)}$.}.
Letting the sum over $i$ in \eqref{eq:decoupling-eq-replica-manifold} run only over $i=1,2$, we recover the decoupling equation \eqref{eq:decoupling-equation-torus-2} for the twist correlator.
Restricting the sum to this range means that we assume $\partial_{w_i} f_\text{cl.} = 0$ for $i > 2$, i.e.~we assume that the result for the partition function on the replica surface does not depend on other moduli of the replica surface than the size of the torus $\tau$ and the length of the entangling interval $z_2 - z_1$.

\begin{figure}
  \centering
  \begin{tikzpicture}
    \begin{scope}[shift={(0,1)},scale=0.7]
      \draw (-5.5,0) node[left] {$\langle (P_p \cO_1 \cO_2 P_p) \cO_3 \cO_4 \rangle \sim $};
      \begin{scope}[shift={(-4,0)}]
        \draw (1,0.8) node[right] {$4$} -- (0.5,0) -- (1,-0.8) node[right] {$3$};
        \draw (0.5,0) -- (-0.5,0);
        \draw (0,0) node[below] {$p$};
        \draw (-1,0.8) node[left] {$1$} -- (-0.5,0) -- (-1,-0.8) node[left] {$2$};
      \end{scope}
      \draw (-1.8,0) node[left] {$\sim$};
      \draw (0,0) ellipse (1.5cm and 1cm);
      \draw (0,1) node[above] {$P_p$};
      \fill (-0.6,0) circle[radius=0.5mm] node[above] {$\cO_1$};
      \fill (0.6,0) circle[radius=0.5mm] node[above] {$\cO_2$};
      \fill (2,0) circle[radius=0.5mm] node[above] {$\cO_3$};
      \fill (3,0) circle[radius=0.5mm] node[above] {$\cO_4$};
    \end{scope}
    \begin{scope}[shift={(0,-1)},scale=0.7]
      \draw (-5.5,0) node[left] {$\langle \cO_1 (P_p \cO_2 \cO_3 P_p) \cO_4 \rangle \sim $};
      \begin{scope}[shift={(-4,0)}]
        \draw (1,0.8) node[right] {$4$} -- (0,0.4) -- (-1,0.8) node[left] {$1$};
        \draw (0,0.4) -- (0,-0.4);
        \draw (0,0) node[right] {$p$};
        \draw (1,-0.8) node[right] {$3$} -- (0,-0.4) -- (-1,-0.8) node[left] {$2$};
      \end{scope}
      \draw (-1.8,0) node[left] {$\sim$};
      \fill (-1,0) circle[radius=0.5mm] node[above] {$\cO_1$};
      \draw (1,0) ellipse (1.5cm and 1cm);
      \draw (1,1) node[above] {$P_p$};
      \fill (0.4,0) circle[radius=0.5mm] node[above] {$\cO_2$};
      \fill (1.6,0) circle[radius=0.5mm] node[above] {$\cO_3$};
      \fill (3,0) circle[radius=0.5mm] node[above] {$\cO_4$};
    \end{scope}
  \end{tikzpicture}
  \caption{Inserting projection operators $P_p$ onto the Verma module of a primary $\cO_p$ into a correlator yields the conformal block with internal weight $h_p$.}
  \label{fig:projection-operators-conformal-block}
\end{figure}

However, as mentioned in the beginning of this section, the admitted monodromy conditions for the decoupling equation \eqref{eq:decoupling-eq-replica-manifold} are more general than those of \eqref{eq:decoupling-equation-torus-2} for the twist correlator.
To see this, recall that conformal blocks of any correlation function can be obtained in two equivalent ways.
Either we can perform OPE contractions between two or more operators and then keep only terms of particular primaries and their descendants in the OPE or equivalently we can insert projection operators onto the Verma modules of these primaries in the correlation function at appropriate places.
The projectors of the latter approach can be thought of as non-local operators acting in a closed line around the operators whose OPE contractions are performed in the former approach (see \figref{fig:projection-operators-conformal-block}).
For the zero-point block on an arbitrary higher genus Riemann surface, there are in general $3(n-1)$ projectors to be inserted corresponding to $3(n-1)$ monodromy conditions.
However, as mentioned above we assume that the partition function on the higher genus Riemann surface depends only on two of the moduli and thus we consider only two of the $3(n-1)$ monodromy conditions.

\begin{figure}
  \centering
  \begin{tikzpicture}
    \begin{scope}[shift={(-3,0)},scale=0.7]
      \draw (0,0) ellipse (3cm and 2cm);
      \draw (-1.5,0) to[out=30,in=150] (1.5,0);
      \draw (-1.65,0.15) to[out=-50,in=230] (1.65,0.15);
      \draw[color=red] ({{2.5*cos(230)}},{{1.5*sin(230)}}) arc(230:-50:2.5cm and 1.5cm) ({{2.5*cos(-50)}},{{1.5*sin(-50)}});
      \fill[color=red] ({{2.5*cos(230)}},{{1.5*sin(230)}}) circle[radius=0.1cm];
      \fill[color=red] ({{2.5*cos(-50)}},{{1.5*sin(-50)}}) circle[radius=0.1cm];
      \draw[color=red] (2.5,0) node[left] {\small{$A$}};
      \draw[->,color=red] (0,1) to[out=120,in=-140] (0.4,1.8);
    \end{scope}
    \begin{scope}[shift={(3,0)},scale=0.7]
      \draw (0,0) ellipse (3cm and 2cm);
      \draw (-1.5,0) to[out=30,in=150] (1.5,0);
      \draw (-1.65,0.15) to[out=-50,in=230] (1.65,0.15);
      \draw[color=red] ellipse (2.55cm and 1.55cm);
      \draw[->,color=red] (0,1.05) to[out=120,in=-140] (0.4,1.85);
      \draw[color=blue] ({{2.45*cos(230)}},{{1.45*sin(230)}}) arc(230:310:2.45cm and 1.45cm) ({{2.45*cos(-50)}},{{1.45*sin(-50)}});
      \fill[color=blue] ({{2.45*cos(230)}},{{1.45*sin(230)}}) circle[radius=0.1cm];
      \fill[color=blue] ({{2.45*cos(-50)}},{{1.45*sin(-50)}}) circle[radius=0.1cm];
      \draw[color=blue] (0,-1.5) node[above] {\small{$A^c$}};
      \draw[->,color=blue] (-0.5,-1) to[out=-140,in=120] (-0.9,-1.8);
    \end{scope}
  \end{tikzpicture}
  \caption{Branch cut structure for large intervals. We can decompose the branch cut along $A$ (denoted in red on the left) into a branch cut along the full spatial circle (denoted in red on the right) and a branch cut in the opposite direction along $A^c$ (denoted in blue on the right).}
  \label{fig:large-interval-branch-cuts}
\end{figure}
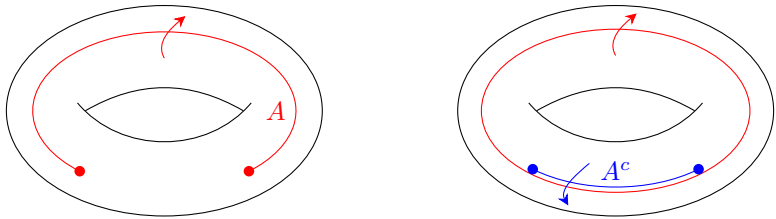

Which monodromy conditions are appropriate for the calculation of the entanglement entropy?
For the conformal block, the monodromy conditions must be taken around the spatial circle and around $z_1,z_2$ \footnote{It is also possible to calculate the modular transformed block for the twist correlator, which we expect to be the dominant contribution at large temperature and small intervals. In this case, the monodromy conditions are taken around $z_1,z_2$ and around the time circle.}.
On the other hand, for the zero-point block on the replica surface the prescription described in this section still leaves open the question of where to put the monodromy conditions -- i.e.~which channel to choose -- in order to obtain the dominant contribution to the partition function from the vacuum block.
Taking the limits of high and low temperature, it is clear that for small intervals one of the monodromy conditions must be taken around the spatial circle for low temperatures and the time circle for high temperatures, while the other monodromy condition must be imposed around the entangling interval $A$ between $z_1$ and $z_2$.
For large intervals, the correct monodromy condition is obtained by reformulating the problem along the lines of \cite{Cardy:2014jwa,Chen:2015kua}.
We separate the branch cut on the torus along $A$ yielding the replica surface into a branch cut along the full spatial circle and a branch cut in the opposite direction along $A^c$ (see \figref{fig:large-interval-branch-cuts}).
We then impose trivial monodromy around $A^c$ to fix the dependence on the size of the entangling interval.
For small temperatures, the monodromy condition around the spatial circle remains unchanged.
However, for high temperatures the monodromy condition around the time circle is now transformed into a monodromy condition around a time circle of size $n\tau$, since the branch cut along the full spatial circle connects all $n$ replica copies together to effectively create a torus with modular parameter $n\tau$.

Note again that it is perfectly valid to use any of the above monodromy conditions for all values of the temperature and entangling interval size.
However, outside of the regimes of validity of the monodromy conditions described above, we don't expect the vacuum block to give the dominant contribution in the semiclassical limit and thus the partition function in this case would be obtained by summing up all of the conformal blocks for different values of the dimensions of the exchanged operators.
The cross-over point between the regimes must be determined by an analysis of these contributions from the exchange of non-identity operators.

Let us also note that explicit calculations for the free fermion case support the arguments presented in this section with regards to the differences between twist correlators and partition functions on $\cR_n$ and with regards to the monodromy condition for large intervals.
Namely, in \cite{Lokhande:2015zma} it was observed that the twist correlator on the torus does not give the correct answer for the entanglement entropy and in particular violates Bose-Fermi equivalence and modular covariance.
This was traced back in \cite{Mukhi:2017rex} to the way in which different spin structures of the replica surface $\cR_n$ combine to give the total answer for the partition function $Z_n$.
The replica surface, being composed of $n$ copies of a torus, has $2n$ nontrivial cycles around which the fermions have either periodic or antiperiodic boundary conditions.
To calculate the total partition function $Z_n$ on the replica surface, it is necessary to sum over all possible spin structures each of which corresponds to a particular choice of boundary conditions around the nontrivial cycles of the replica surface.
For small intervals, the replica surface essentially factorizes into $n$ unconnected torus copies.
In this limit, it was shown in \cite{Mukhi:2017rex} that $Z_n$ is given by an ``uncorrelated'' sum where the summation over spin structures is performed for each of the $n$ tori separately.
For large intervals, $Z_n$ is given by a ``correlated'' sum where only one sum over spin structures is performed, i.e.~we take equal boundary conditions around the time resp.~space circles of each of the $n$ replica copies \cite{Mukhi:2017rex}.
In this case, the partition function $Z_n$ of the replica surface is essentially given by the partition function on a torus with modular parameter $n\tau$.

\section{Entanglement entropy at large central charge}
\label{sec:EE-thermal-states}

In this section, we present the calculation of the entanglement entropy on the torus at large central charge.
As explained in the previous section, the entanglement entropy can be obtained from the partition function $Z_n$ on the replica surface $\cR_n$ which decomposes into zero-point conformal blocks.
The claim we want to investigate is that at large central charge $c \to \infty$ and for $n \to 1$ the dominant contribution to $Z_n$ comes from the vacuum block with $h_p = h_q = 0$.
The derivation of this statement proceeds as follows.
First, it is necessary to show that the semiclassical limit is well-defined not only for $h_{p,q} = \cO(c)$ but also for $h_{p,q} = \cO(c^0)$.
This means that for $h_{p,q} = \gamma c$ the limits $\lim_{c \to \infty}$ and $\lim_{\gamma \to 0}$ of the conformal block commute.
A discussion of this point starting from the recursion relation for torus conformal blocks is relegated to app.~\ref{app:recursion-relations}.
In the next step we solve the decoupling equation \eqref{eq:decoupling-equation-torus-2} perturbatively in $\varepsilon = n-1$ up to first order.
Imposing the monodromy conditions derived in \secref{sec:monodromy-method-torus} then yields the conformal block from which we finally extract the entanglement entropy.

We first consider the limits of high and low temperature as well as small and large entangling interval size in \secref{sec:EE-limits}.
Each combination of these limits comes with different monodromy conditions as explained in the previous section.
We find agreement with the known universal results in the limits where the torus degenerates into a cylinder.
We then examine the conditions on the CFT spectrum and OPE coefficients under which these results extend to intermediate temperature and interval size regimes in \secref{sec:holographic-cfts}, leading to the conclusion that for holographic CFTs the results from \secref{sec:EE-limits} are valid for all temperatures and interval sizes.
Moreover, we consider the case of multiple intervals on the torus.
Finally, we numerically check the assumption on the dominance of the vacuum block in \secref{sec:vacuum-block-dominance}.

\subsection{Limiting cases} 
\label{sec:EE-limits}

\subsubsection{Low temperature and small intervals}
In the low temperature limit $\beta \to \infty$ the torus degenerates into a cylinder with periodic space direction.
For the cylinder, the entanglement entropy of a single interval can be obtained directly by mapping this cylinder to the plane and using the known formula for the entanglement entropy on the plane \cite{Calabrese:2004eu},
\begin{equation}
  S_A = \frac{c}{3} \log\left(\sin(\pi(z_2 - z_1))\right) + \text{const.}
  \label{eq:EE-low-temperature}
\end{equation}

To obtain the same result from the monodromy method, we series expand $\Psi_{pq}$ and $f_\text{cl.}$ in $n-1$: $\Psi_{pq} = \sum_k \Psi_{pq}^{(n)} (n-1)^k$ and $f_\text{cl.} = \sum_k f_n (n-1)^k$.
The decoupling equation \eqref{eq:decoupling-equation-torus-2} at zeroth order in $n-1$ becomes
\begin{equation}
  [\partial_z^2 - 2\pi i \partial_\tau f_0]\Psi_{pq}^{(0)}(z) = 0.
\end{equation}
This is solved by
\begin{equation}
  \Psi^{(0)}_{pq}(z) = \exp(\pm \sqrt{2\pi i \partial_\tau f_0} z).
\end{equation}
By a coordinate transformation $u = \exp(-2\pi i z)$, we obtain $\tilde \Psi_{pq}^{(0)}(u) = u^{h_\Psi} \Psi_{pq}^{(0)}(z = i \log(u)/(2\pi))$.
Imposing trivial monodromy of $\tilde \Psi_{pq}^{(0)}(u)$ around $u=0$ is equivalent to antiperiodic monodromy conditions for $\Psi_{pq}^{(0)}(z)$ around the spatial circle of the torus, $\Psi_{pq}^{(0)}(z+1) = -\Psi_{pq}^{(0)}(z)$.
As expected, this implies that $f_0$ is equal to the leading order in $c$ of the vacuum character on the torus,
\begin{equation}
  f_0 = \pi i \tau/2 = -\beta/4 \hspace{1cm} \Leftrightarrow e^{-c/6 f_0} = e^{c/24 \beta} = \left.\chi_{h=0}(\beta)\right|_{c\to\infty}.
\end{equation}
At first order in $n-1$, the decoupling equation is given by
\begin{equation}
  [\partial_z^2 - 2\pi i \partial_\tau f_0]\Psi_{pq}^{(1)}(z) + m(z) \Psi_{pq}^{(0)}(z) = 0,
\end{equation}
yielding
\begin{equation}
  \Psi_{pq}^{(1)}(z) = \frac{e^{-i\pi z}}{2\pi i}\int^z dx\, m(x) e^{i\pi x}\Psi_{pq}^{(0)}(x) - \frac{e^{i\pi z}}{2\pi i} \int^z dx\, m(x) e^{-i\pi x}\Psi_{pq}^{(0)}(x),
\end{equation}
where $m(z)$ is given by
\begin{equation}
  m(z) = \sum_i \left(\frac{1}{2}(\wp(z-z_i) + 2\eta_1) + (-1)^{i+1}(\zeta(z-z_i) + 2\eta_1 z_i)\partial_{z_2} f_1\right) - 2\pi i \partial_\tau f_1.
\end{equation}
To compute the conformal block we impose trivial monodromy around $z_1,z_2$ which is equivalent to the vanishing of
\begin{equation}
  \oint_{z_1,z_2} dx\, m(x) e^{\pm i\pi x} \Psi_{pq}^{(0)}(x).
\end{equation}
This gives the $z_1,z_2$ dependence of $f_1$,
\begin{equation}
  f_1 = \log(\sin(\pi(z_2-z_1))) + C_1(\tau).
\end{equation}
From trivial monodromy around $u=0$ we find the $\tau$ dependence to be $\partial_\tau f_1 = 0$, which implies that $C_1(\tau) = \text{const.}$ is independent of $\tau$.
The antiholomorphic conformal block $\bar f_1$ gives the same result as the holomorphic one.
Then the entanglement entropy is given by $S_A = \frac{c}{6} (f_1 + \bar f_1)$, in agreement with \eqref{eq:EE-low-temperature}.
In this limit, the OPE vacuum block of the twist correlator gives the same results, since it is given by the same monodromy method as the zero-point vacuum block of the replica partition function computed in this section.

\subsubsection{Low temperature and large intervals}
In this limit, we demand trivial monodromy around the spatial circle (i.e.~around $u=e^{-2\pi i z}=0$) as well as trivial monodromy around $A^c$ (i.e.~around $z_1,z_2-1$).
This gives the same vacuum block and thus the same entanglement entropy \eqref{eq:EE-low-temperature} as in the small interval case at low temperature.
Also in this case, the twist correlator gives the correct result.

\subsubsection{High temperature and small intervals}
In the high temperature limit $\beta \to 0$ the torus again degenerates into a cylinder, now with periodic time direction.
As in the low temperature case, the entanglement entropy of a single interval can be obtained by mapping this cylinder to the plane \cite{Calabrese:2004eu},
\begin{equation}
  S_A = \frac{c}{3} \log\left(\frac{\tau}{i\pi}\sinh\left(\frac{i\pi}{\tau}(z_2 - z_1)\right)\right) + \text{const.}
  \label{eq:EE-high-temperature}
\end{equation}

In the monodromy method, we impose trivial monodromy around the time circle and around $z_1,z_2$.
As in the low temperature limit, we solve the decoupling equation \eqref{eq:decoupling-equation-torus-2} in a series expansion around $n-1$.
An analogous calculation as above yields
\begin{equation}
  f_\text{cl.} = -\frac{\pi i}{2\tau} + \epsilon\log\left(\tau\sinh\left(\frac{\pi i}{\tau}(z_2-z_1)\right)\right) + \text{const.} = f_0 + \epsilon f_1
  \label{eq:conformal-block-high-temperature-small-intervals}
\end{equation}
At zeroth order in $\epsilon$ we recover the leading order in $c$ of the vacuum character $\chi_{h=0}(\beta) = e^{\frac{c}{24} \frac{4\pi^2}{\beta}} = e^{-\frac{c}{6}(-\frac{\pi i}{2\tau})}$.
The entanglement entropy given by the first order contribution, $S_A = \frac{c}{6} (f_1 + \bar f_1)$, agrees with the result from the cylinder \eqref{eq:EE-high-temperature}.

From the twist correlator point of view, we obtain the high temperature limit by a modular transformation $\tau \to -1/\tau$ from the low temperature result.
The two point function on the torus transforms covariantly,
\begin{equation}
  \langle \cO_1(-z_1/\tau,-\bar z_1/\bar\tau) \cO_2(-z_2/\tau,-\bar z_2/\bar\tau) \rangle_{-1/\tau} = (-\tau)^{h_1+h_2}(-\bar\tau)^{\bar h_1+\bar h_2} \langle \cO_1(z_1,\bar z_1) \cO_2(z_2,\bar z_2) \rangle_\tau.
\end{equation}
If the two point function of twist operators is dominated by a single conformal block, the conformal block for the modular transformed $\tau$ is obtained from the modular transformation properties of the correlation function as
\begin{equation}
  f_\text{cl.}^{-1/\tau}(-z_1/\tau,-z_2/\tau) = -b^2(h_1 + h_2)\log(-\tau) + f_\text{cl.}^\tau(z_1,z_2).
\end{equation}
Therefore, we can immediately read off the high temperature behavior of the twist correlator from
\begin{equation}
  \left.f_\text{cl.}^\tau(z_1,z_2)\right|_{\tau \to \infty} = \left[f_\text{cl.}^{-1/\tau}(-z_1/\tau,-z_2/\tau) + b^2(h_1 + h_2) \log(-\tau)\right]_{\tau \to \infty}
\end{equation}
For $h_1 = h_2 = 1/2\epsilon$, $\left.f_\text{cl.}^\tau(z_1,z_2)\right|_{\tau \to 0} = \frac{\pi i\tau}{2} + \epsilon \log(\sin(\pi(z_2-z_1))) + \text{const.}$ we again obtain \eqref{eq:conformal-block-high-temperature-small-intervals}.
Thus the twist correlator agrees with the replica partition in this limit.

Let us note that is also possible to determine the high temperature expansion of twist correlator by applying the modular transformation to the monodromy conditions.
At low temperatures, we demand trivial monodromy around the spatial circle of the torus at time.
Since the modular transformation $\tau \to -1/\tau$ exchanges the time and space directions of the torus, we can obtain the high temperature behavior of the twist correlator by demanding trivial monodromy around the time circle of the torus, showing directly the equivalence between the twist correlator result and the replica partition function in the limit of high temperature and small entangling intervals.

\subsubsection{High temperature and large intervals}
As explained in \secref{sec:monodromy-method-replica-manifold}, the correct monodromy condition of the zero-point block on the replica surface for large intervals imposes trivial monodromy around $A^c$ and $z \to z + n\tau$.
To zeroth order in $n-1$ the decoupling equation \eqref{eq:decoupling-equation-torus-2} is solved by $\Psi_{pq}^{(0)}(z) = \exp(\pm\sqrt{2\pi i\partial_\tau f_0})$.
Imposing $\Psi_{pq}^{(0)}(z + n\tau) = - \Psi_{pq}^{(0)}(z)$ yields
\begin{equation}
  f_0 = - \frac{\pi i}{2n^2\tau}.
\end{equation}
To first order, the solution reads
\begin{equation}
  \Psi_{pq}^{(1)}(z) = n\tau \frac{e^{-\frac{i\pi z}{n\tau}}}{2\pi i}\int^z dx\, m(x) e^{\frac{i\pi x}{n\tau}}\Psi_{pq}^{(0)}(x) - n\tau\frac{e^{\frac{i\pi z}{n\tau}}}{2\pi i} \int^z dx\, m(x) e^{-\frac{i\pi x}{n\tau}}\Psi_{pq}^{(0)}(x).
\end{equation}
Imposing trivial monodromy around $A^c$ and expanding in $n-1$ we obtain
\begin{equation}
  f_\text{cl.} = -\frac{\pi i}{2\tau} + (n-1)\left(\frac{i\pi}{\tau} + \log\left(\tau\sinh\left(\frac{i\pi}{\tau}(1 - (z_2 - z_1))\right)\right)\right) + \text{const.}
  \label{eq:conformal-block-high-temperature-large-intervals}
\end{equation}
giving
\begin{equation}
  S_A = \frac{c}{3} \left(\frac{i\pi}{\tau} + \log\left(\frac{\tau}{i\pi}\sinh\left(\frac{i\pi}{\tau}(1 - (z_2 - z_1))\right)\right)\right) + \text{const.}
  \label{eq:EE-high-temperature-large-intervals}
\end{equation}
As expected from general arguments \cite{Azeyanagi:2007bj,Cardy:2014jwa}, the difference between the entanglement entropy for $A$ and for the complement $A^c$ in the limit of large interval size is given by the thermal entropy $S(\beta) = \beta^2 (-\beta^{-1}\partial_\beta \log Z(\beta)) = \frac{c}{3} \frac{2\pi^2}{\beta} = \frac{c}{3} \frac{i\pi}{\tau}$, using the partition function $Z(\beta \to 0) = \exp(\frac{c}{12}\frac{4\pi^2}{\beta})$ from the Cardy formula.
The twist correlator, on the other hand, cannot reproduce this feature as is easy to see by applying the modular transformation argument from the last section which gives \eqref{eq:conformal-block-high-temperature-small-intervals} in disagreement with \eqref{eq:conformal-block-high-temperature-large-intervals}.

\subsection{Holographic CFTs}
\label{sec:holographic-cfts}
We now argue that the results in the limits considered in the previous section are valid for all temperatures and interval sizes in holographic CFTs.
This statement holds if the vacuum block computed in the previous section gives the dominant contribution to the partition function on the replica manifold $\cR_n$ in the large central charge limit.

Let us first consider the case $n=1$, i.e.~the zeroth order in the $n-1$ expansion.
In this order, the monodromy method calculates the vacuum character to leading order in $c$.
If this vacuum character dominates the partition function, $Z(\beta)$ takes on the universal form
\begin{equation}
  Z(\beta) = \left\{
    \begin{array}{ll}
      \exp\bigl(\frac{c}{12}\beta\bigr) &,~ \beta > 2\pi\\
      \exp\bigl(\frac{c}{12}\frac{4\pi^2}{\beta}\bigr) &,~ \beta < 2\pi
    \end{array}
  \right.
  \label{eq:universal-partition-function}
\end{equation}
for all temperatures.
For consistency of the computation method, dominance of the conformal block to first order in $n-1$ also requires dominance of the zeroth order in $n-1$.
It has been argued in \cite{Hartman:2014oaa} that a partition function of the form \eqref{eq:universal-partition-function} is characteristic of a holographic CFT.
Thus it is a necessary condition that the CFT in question be holographic in order for the results of \secref{sec:EE-limits} to hold at arbitrary temperatures.

More explicit conditions on the CFT in question can be given following an argument for dominance of the vacuum block in the zero temperature case which goes as follows \cite{Hartman:2013mia}.
The arguments below hold for CFTs with OPE coefficients $C^p_{21}$ that grow at most exponentially with $c$ \footnote{This is equivalent to the requirement that correlation functions obey the cluster decomposition principle and are smooth in a neighborhood of the point where multiple operator insertion points coincide \cite{Hartman:2013mia}.} and a density of states for light operators that does not grow with $c$.
Here light operators mean operators with conformal weight $h,\bar h < h_\text{gap}$ where $h_\text{gap}$ is of the order of the central charge.
In the large central charge limit, the conformal block expansion in the $s$-channel of the four-point twist correlator on the plane takes on the form
\begin{equation}
  \langle \sigma_n(0) \sigma_n(x) \sigma_n(1) \sigma_n(\infty) \rangle = \sum_p C^p_{21} \exp\left(-\frac{c}{6}\bigl(f_\text{cl.}(h_{\sigma_n}/c,h_p/c,x) + \bar f_\text{cl.}(\bar h_{\sigma_n}/c,\bar h_p/c,\bar x)\bigr)\right).
  \label{eq:conformal-block-expansion-4pt-plane}
\end{equation}
On account of the cluster decomposition principle, the vacuum exchange is the leading contribution in the $s$-channel around $x=0$ and in the $t$-channel around $x=1$.
This implies that the contribution of the heavy operators with $h,\bar h > h_\text{gap}$ in the $s$-channel in some finite region around $x=0$ is suppressed exponentially and similarly for the $t$-channel around $x=1$.
The sparse spectrum of the light operators allows ignoring the multiplicity factors for $h,\bar h < h_\text{gap}$.
Consider first the scenario that the OPE coefficients grow subexponentially with $c$.
In this case, we can also ignore the coefficient $C^p_{21}$ and the sum over the light operators in \eqref{eq:conformal-block-expansion-4pt-plane} is dominated exponentially in $e^{-c}$ by its largest term.
If the semiclassical conformal block $f_\text{cl.}$ is a monotonically increasing function with $h_p/c$, this implies that the vacuum block with the lowest possible $h_p/c = 0$ dominates.
In the case that the OPE coefficients grow exponentially with $c$, the sum over the light operators gives
\begin{equation}
  \int_0^{h_\text{gap}} dh_p d\bar h_p \exp\left(\frac{c}{6}\bigl(g(h_p/c,\bar h_p/c) - f_\text{cl.}(h_{\sigma_n}/c,h_p/c,x) - \bar f_\text{cl.}(\bar h_{\sigma_n}/c,\bar h_p/c,\bar x)\bigr)\right),
  \label{eq:bound-vacuum-block-dominance}
\end{equation}
where $g(h_p/c,\bar h_p/c)$ contains the contribution of the OPE coefficients and multiplicities.
This integral is dominated either by the endpoints of the integration or by a saddle point.
Near coincident points, the leading universal term in any correlation function is given by the vacuum exchange.
Therefore, in a finite region around $x=0$ and $x=1$, the integral in \eqref{eq:bound-vacuum-block-dominance} is dominated by the endpoint at $h_p=0$.
However, now -- unlike for subexponentially growing OPE coefficients -- it is not possible to exclude saddle points dominating the integral \eqref{eq:bound-vacuum-block-dominance} in some finite subset of $x \in [0,1]$.
Examples of such saddle points have been found in \cite{Belin:2017nze} for genus two partition functions computing the third Rényi entropy on the plane.

In summary, the vacuum block dominates correlation functions in a finite region around coincident operator insertion points if the conformal block $f_\text{cl.}$ increases monotonically with the internal conformal weights $h_{p,q}$.
In \secref{sec:vacuum-block-dominance}, we present numerical evidence that this is indeed the case.
Therefore, the results of \secref{sec:EE-limits} are valid for all temperatures and interval sizes assuming subexponentially growing OPE coefficients and a sparse spectrum of light operators.
For exponentially growing OPE coefficients, the results are still valid in a finite region around the respective limiting points.
The points at which the different limits exchange dominance are obtained as follows.
From the requirement of consistency of our results at $n=1$ with the partition function \eqref{eq:universal-partition-function} we see that the low and high temperature regimes exchange dominance at the Hawking-Page phase transition point $\beta = 2\pi$.
The transition between small and large interval behavior in the high temperature regime is estimated by equating the results for the conformal block in the small and large interval regime.
The dominant contribution comes from the smaller conformal block.

Assuming the aforementioned restrictions on the CFT spectrum and OPE coefficients, the results match perfectly with the predictions from the RT formula \cite{Ryu:2006bv,Azeyanagi:2007bj}.
There is also a direct way to implement the monodromy computation for the calculation of the entanglement entropy in the dual gravity theory, as was shown in \cite{Faulkner:2013yia} at zero temperature.
This was generalized to the finite temperature case in \cite{Barrella:2013wja,Chen:2015kua}.
We obtain the same monodromy method as \cite{Barrella:2013wja,Chen:2015kua} from the CFT side.
This clearly shows that our results are valid for holographic CFTs.

Furthermore, the high temperature and large interval size result \eqref{eq:conformal-block-high-temperature-large-intervals} agrees with a CFT calculation of the vacuum sector contribution to the entanglement entropy done in \cite{Chen:2015kua} using complementary techniques to obtain the vacuum block.
Specifically, the conformal block is obtained in \cite{Chen:2015kua} by an explicit construction of the Virasoro generators and descendant states in the twisted sector of a $\ZZ_n$ orbifold.
This allows for a series expansion of the Rényi entropy.
Moreover, in the limit $n \to 1$, the authors of \cite{Chen:2015kua} find that the leading order of the semiclassical vacuum block is given by an expression in terms of the four-point function of twist-operators on the plane, which gives the same conformal block \eqref{eq:conformal-block-high-temperature-large-intervals} that we obtain using the monodromy method.

Let us also note that the universal form of the partition function \eqref{eq:universal-partition-function} in holographic theories can be derived from the same vacuum block dominance argument as the universal form of the entanglement entropy.
The partition function can be expanded in zero-point blocks on the torus either in a low temperature expansion (zero-point blocks are Virasoro characters) or in a high temperature expansion (zero-point blocks are modular transformed Virasoro characters).
From the known form of the Virasoro characters, the leading order contribution in the central charge of the characters is given by $\chi \sim Q^{h_q - c/24}$ where $Q = e^{2\pi i \tau}$ resp.~$Q = e^{-2\pi i/\tau}$ in the low and high temperature expansions.
Since $(-6/c)\log \chi$ is an increasing function of $h_q$, the same arguments as for the entanglement entropy given above apply to the partition function which is dominated by the vacuum character with $h_q = 0$.

\subsection{Multiple intervals}
\label{sec:multiple-intervals}
The generalization to an entangling interval consisting of the union of multiple intervals, $A = [z_1,z_2] \cup [z_3,z_4] ...  [z_{2N-1},z_{2N}]$, is straightforward.
The decoupling equation is given by
\begin{equation}
  \left[\partial_z^2 + \sum_{i=1}^{2N} \left(\frac{1}{4}(n-1/n)(\wp(z-z_i) + 2\eta_1) - \partial_{z_i} f_\text{cl.}(\zeta(z-z_i) + 2\eta_1 z_i)\right) - 2\pi i\partial_\tau f_\text{cl.}\right] \Psi(z) = 0.
\end{equation}
We impose trivial monodromy around $N$ pairs $(i,j)$ of interval endpoints $z_i,z_j$ to fix the $\partial_{z_i} f_\text{cl.}$.
The temperature dependence is fixed by demanding trivial monodromy around either the spatial circle, a time circle of size $\tau$ or a time circle of size $n\tau$ depending on the temperature and total entangling interval size $|A| = \sum_i |z_{2i} - z_{2i-1}|$.
This yields
\begin{itemize}
\item low temperature: trivial monodromy for $z \to z + 1$
  \begin{equation}
    S_A = \frac{c}{3} \sum_{(i,j)} \log\left(\sin(\pi(z_i - z_j))\right) + \text{const.}    
  \end{equation}
\item high temperature and small total interval size: trivial monodromy for $z \to z + \tau$
  \begin{equation}
    S_A = \frac{c}{3} \sum_{(i,j)} \log\left(\frac{\tau}{i\pi}\sinh\left(\frac{i\pi}{\tau}(z_i - z_j)\right)\right) + \text{const.}
  \end{equation}
\item high temperature and large total interval size: trivial monodromy for $z \to z + n\tau$
  \begin{equation}
    S_A = \frac{c}{3} \left(\frac{i\pi}{\tau} + \sum_{(i,j)} \log\left(\frac{\tau}{i\pi}\sinh\left(\frac{i\pi}{\tau}(z_i - z_j)\right)\right)\right) + \text{const.}
  \end{equation}
\end{itemize}
Which monodromy condition and which combination of pairs $(i,j)$ to take, i.e.~in which channel to expand the conformal block, depends on the interval size.
We expect the dominant contribution to come from the channel with the smallest $f_\text{cl.}$, in which case we find agreement with the RT formula.
However, we caution that this argument depends on the vacuum block dominating the partition function, which we have checked only for a single interval.

One particular interesting special case of the above calculation is the time dependence of the entanglement entropy between two intervals on opposite boundaries of a two-sided BTZ black hole.
The two-sided BTZ black hole is dual to the thermofield double state.
On the CFT side, the entanglement entropy is obtained by positioning the two intervals at a distance $\tau/2$ in the time coordinate on the torus.
To obtain the time evolution of the entanglement entropy, it is necessary to perform an analytic continuation of the endpoints \cite{Hartman:2013qma},
\begin{equation}
  \begin{aligned}
    z_1 = \bar z_1 = 0 && z_2 = \bar z_2 = L\\
    z_3 = 2t + L + \tau/2, \bar z_3 = -2t + L + \bar\tau/2 &\hspace{1cm}& z_4 = 2t + \tau/2, \bar z_4 = -2t + \bar\tau/2,
  \end{aligned}
\end{equation}
where $L$ is the size of the entangling interval taken to be equal on both boundaries and $t$ is the time coordinate at which both parts of the entangling interval are placed on the asymptotic boundaries of the wormhole\footnote{Note that this time coordinate has nothing to do with the euclidean time coordinate on the torus on which we calculate the finite temperature correlator in the euclidean CFT.} (see \cite{Hartman:2013qma} for more details on the setup).
For small interval size $L$, there are two conformal blocks to consider.
At early times, i.e.~for small $t$, the dominant contribution comes from imposing trivial monodromy around $z_1,z_4$ and $z_2,z_3$, while at late times the vacuum block with trivial monodromy around $z_1,z_2$ and $z_3,z_4$ dominates.
Taking into account that due to the analytic continuation $f_\text{cl.} \neq \bar f_\text{cl.}$, the corresponding entanglement entropy is given by
\begin{equation}
  S_A = \frac{2c}{3} \log\left(\frac{\tau}{i\pi}\cosh\left(\frac{2\pi i}{\tau}t\right)\right) + \text{const.}
\end{equation}
at early times and by
\begin{equation}
  S_A = \frac{2c}{3} \log\left(\frac{\tau}{i\pi}\sinh\left(\frac{i\pi}{\tau}L\right)\right) + \text{const.}
\end{equation}
at late times.
This reproduces the phase transition in the dual RT surfaces from geodesics that connect the two boundaries through the interior of the two-sided black hole at early times to disconnected geodesics on opposite boundaries that do not enter the black hole interior at late times \cite{Hartman:2013qma}.

\subsection{Vacuum block dominance}
\label{sec:vacuum-block-dominance}

In this section, we provide numerical evidence that the vacuum block exponentially dominates the twist correlator in the large $c$ limit.
For simplicity, we restrict to the single interval case.
Assuming the same conditions on the spectrum and OPE coefficients of the CFT detailed in the last section, we need to show that the conformal block monotonically increases with the weight of the internal operators $h_{p,q}$.

For the zero temperature case, this was done numerically in \cite{Hartman:2013mia} for arbitrary $n$, giving evidence that the Rényi entropies are given by the vacuum conformal block contribution only.
However, the calculation is much simpler if we restrict to $n$ close to one which implies $h_p/c \gg h_i/c \to 0$ for $i=1,2,3,4$.
In this limit, the conformal block can be obtained in closed form from the monodromy calculation by a WKB expansion in $1/(h_p/c)$ \cite{Zamolodchikov1987},
\begin{equation}
  \frac{c}{6} f_\text{cl.}(0,h_p/c,x) = h_p \left(\pi \frac{K(1-x)}{K(x)} - \log 16\right),
\end{equation}
where $K(x)$ is the complete elliptic integral of the first kind.
Thus, $f_\text{cl.}$ is an increasing function of $h_p/c$ if $\pi K(1-x)/K(x) - \log 16 > 0$ which can easily be checked to be fulfilled for $x < 1/2$.
In fact, $f_\text{cl.}(0,h_p/c,x) - f_\text{cl.}(0,h_p/c,1-x)$ reaches its crossover point exactly at $x = 1/2$, confirming that at this point dominance is exchanged from the $s$ to the $t$-channel block.

Applying the same arguments as on the plane to the case of the torus, it is clear that the vacuum block dominates if the semiclassical block is an increasing function of $h_p$ and $h_q$.
Without loss of generality, we take $z_1=0$ in the following.
Restricting again to $n \approx 1$, we need to find the semiclassical block in the limit $h_{p,q}/c \gg h_{1,2}/c \to 0$.
Unlike on the plane, however, $f_\text{cl.}$ can not easily be obtained in a closed form expression from the monodromy calculation in this limit\footnote{The reason for this is that the solution of the decoupling equation on the torus takes on the schematic form of an integral over $\sqrt{A \partial_{z_2} f_\text{cl.} + B \partial_\tau f_\text{cl.}}$ for some functions $A$ and $B$, from which it is not easily possible to extract $\partial_{z_2} f_\text{cl.}$ and $\partial_\tau f_\text{cl.}$. The solution of the decoupling equation on the plane, on the other hand, is given by an integral over $\sqrt{A \partial_x f_\text{cl.}}$ from which $\partial_x f_\text{cl.}$ can be factored out immediately.}.
Thus, we apply a series expansion in $y = e^{-2\pi i z_2}-1$ and $Q = e^{2\pi i \tau}$ on top of the WKB approximation in $1/(h_{p,q}/c)$.
While this yields a very precise numerical approximation to the true value of the conformal block if enough terms are included in the series, the expansion has a restricted domain of validity in the $y,Q$ plane.
In particular, the series expansion in the cross-ratio $x$ of the four-point block on the plane converges for $|x|<1$ \cite{Zamolodchikov1987}, therefore we expect the series expansion of the two-point block on the torus to have a convergence radius of $|y|=1$ (the torus block reduces to the block on the plane in the limit $Q \to 0$).
The numerics confirm this expectation.
For $|y| > 1$, we observe large fluctuations in the value of the conformal block as we include more terms in the series expansion.
The numerics for the convergence radius in $Q$ is less clear, but also in this variable we observe large fluctuations close to $Q = e^{-2\pi}$.
Thus, we can check the vacuum block dominance only in a restricted region around the origin in $y$, corresponding to small intervals.
The restricted convergence radius in $Q$ is not limiting since above the Hawking-Page transition temperature given by $Q=e^{-2\pi}$, we expect the block in the high temperature expansion to dominate.
The high temperature expansion of the conformal block is given by a series expansion in $e^{2\pi i z_2/\tau}-1$ and $e^{-2\pi i/\tau}$.
In the $h_{p,q}/c \gg h_{1,2}/c$ limit, the series coefficients are equal to those of the low temperature expansion.
Moreover, in the same limit at high temperatures and for $n \to 1$ the conformal block in the large interval limit (where the monodromy condition is taken around a time circle of size $n\tau$) is equivalent to the conformal block in the small interval limit with the replacement $z_2 \to 1-z_2$.

\begin{figure}
  \centering
  \includegraphics[width=.48\textwidth]{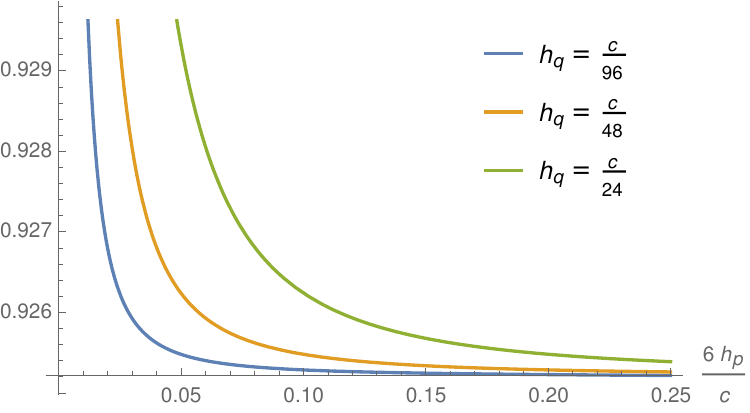}
  \includegraphics[width=.48\textwidth]{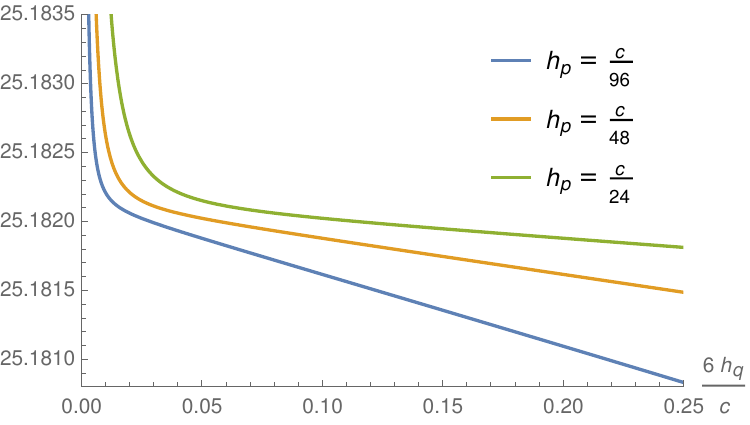}
  \caption{Derivative of the semiclassical conformal block $f_\text{cl.}$ w.r.t.~$h_p$ (left) and $h_q$ (right) for $z_2 = 0.1$, $\beta = 4\pi$ and different values of $h_p$ and $h_q$ in the range $[0,c/24]$. Note that the plotted value is greater than zero in all cases, showing that the semiclassical conformal block increases with $h_p$ and $h_q$. The series expansion used in this figure was truncated at order 10 in both $y$ and $Q$.}
  \label{fig:vacuum-block-dominance-small-temperatures}
\end{figure}

We show some plots of $\frac{\partial f_\text{cl.}}{\partial h_p}$ and $\frac{\partial f_\text{cl.}}{\partial h_q}$ in \figref{fig:vacuum-block-dominance-small-temperatures} for small temperatures and values of $y$ and $Q$ inside the convergence radius.
The plots for the conformal block in the high temperature limit show no significant differences from the ones in the low temperature limit.
We find in all cases that inside the convergence radius of the series expansion $\frac{\partial f_\text{cl.}}{\partial h_{p,q}} > 0$, i.e.~the assumption of vacuum block dominance is fulfilled

\begin{figure}
  \centering
  \includegraphics[width=.7\textwidth]{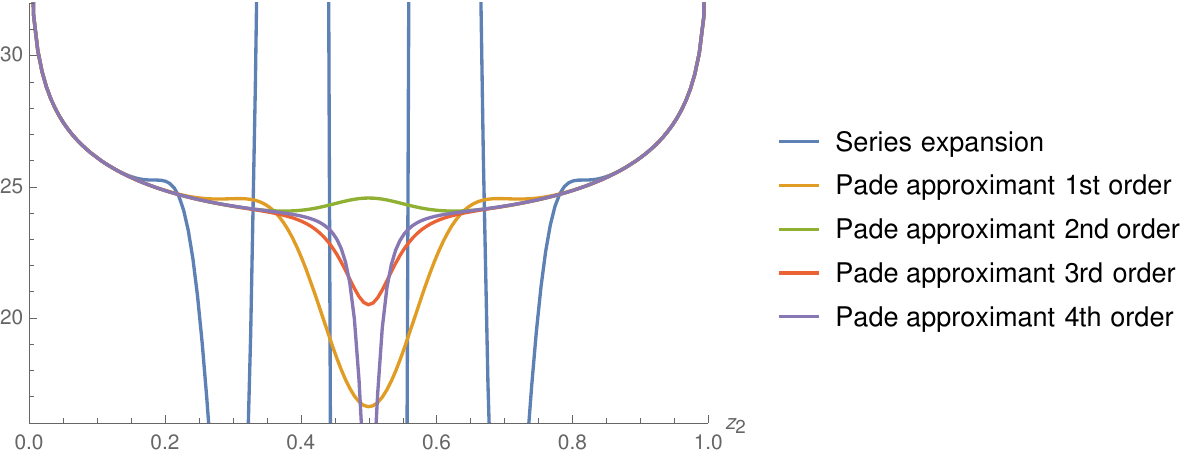}\\
  \includegraphics[width=.7\textwidth]{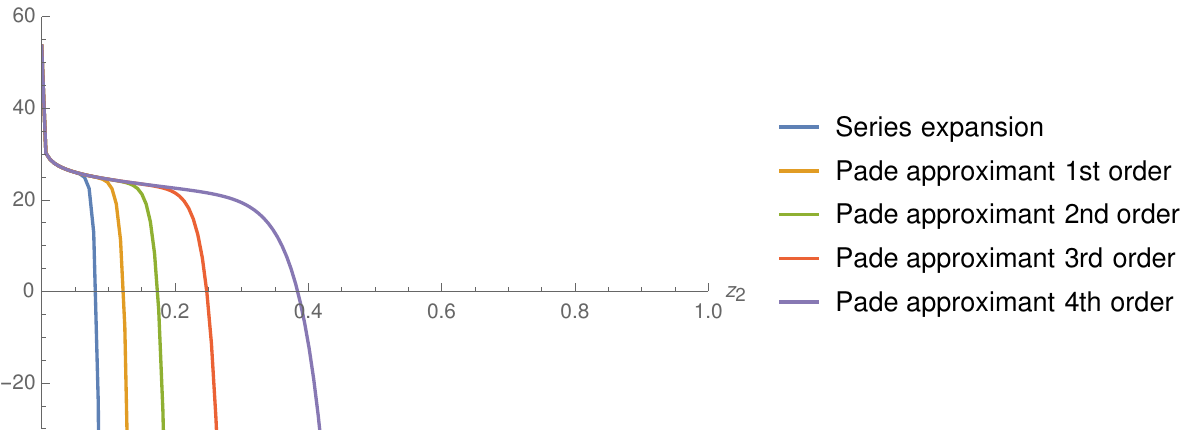}  
  \caption{Derivative of the Padé approximation of the semiclassical conformal block w.r.t.~the internal conformal weight plotted against $z_2$. The upper plot shows the conformal block at $\beta = 4\pi$ in the low temperature expansion and the lower plot the block at $\beta = \pi$ in the high temperature expansion. For plotting convenience we set $h_p = h_q$ which gives a derivative of $f_\text{cl.}$ that is constant for all $h_p$. As we increase the order of the denominator polynomial in the Padé approximation, the fluctuations in $f_\text{cl.}$ outside of the convergence radius of the series expansion decrease. Moreover, multiple Padé approximants converge to the same value outside of the convergence radius. The convergence radius $|y|=1$ is reached at $z_2 = 1/6$ in the upper plot and $z_2 = \frac{\log(2)}{4\pi} = 0.0552$ in the lower plot. The series expansion approximated by the Padé approximants in this figure was truncated at order 10 in both $y$ and $Q$.}
  \label{fig:vacuum-block-dominance-pade-approximation}
\end{figure}

While it is not possible to find an analytic continuation for the conformal block from a truncated series expansion, we can use a Padé approximant to get a heuristic approximation of the series outside its convergence radius.
The Padé approximation works by replacing the truncated series expansion by a rational function whose Taylor expansion agrees with the series expansion up to the order in which the truncation was performed \cite{BakerGraves-Morris}.
In many cases, this approximation has a better radius of convergence than the original series expansion due to poles in the function limiting the radius of convergence of its Taylor expansion being taken into account in the Padé approximation.
We plot the Padé approximant of $f_\text{cl.}$ in the special case $h_p = h_q$ for different orders of the denominator polynomial in \figref{fig:vacuum-block-dominance-pade-approximation} depending on $z_2$ (the size of the entangling interval).
We find that different Padé approximants for $f_\text{cl.}$ converge to the same value and yield $\frac{\partial f_\text{cl.}}{\partial h_{p,q}} > 0$ in a finite region outside of the convergence radius $|y|=1$ of the series expansion of $f_\text{cl.}$.
While this is not a formal proof, it does indicate that vacuum block dominance holds also outside of the convergence radius.

\section{Discussion}
\label{sec:discussion}
Let us briefly summarize the main points investigated in this publication.
In the first part in \secref{sec:monodromy-methods}, we took a detailed look at monodromy methods for the computation of arbitrary conformal blocks on the torus as well as zero-point conformal blocks on the special higher genus Riemann surface relevant to the calculation of the entanglement entropy via the replica trick.
We then applied the derived monodromy methods to the calculation of the entanglement entropy in \secref{sec:EE-thermal-states}.
We found that for holographic CFTs, the vacuum conformal block dominates the partition function on the replica surface and therefore the entanglement entropy takes on a universal form in agreement with the RT formula.

The main advancement compared to previous work on entanglement entropy for finite size and finite temperature systems at large central charge is as follows.

First of all, we clarified the role of the correlation function of twist operators on the torus compared to the partition function $Z_n$ on the replica surface $\cR_n$.
The twist correlator only gives the correct results for small intervals or low temperature, while the replica partition function $Z_n$ must give the entanglement entropy in all cases by construction.
The reason why the twist correlator agrees with $Z_n$ for small intervals or low temperature is evident from the monodromy method.
The conformal blocks of the twist correlator obey the same monodromy method as the zero-point conformal blocks for $Z_n$ apart from the set of allowed monodromy conditions, which for the twist correlator blocks is a subset of those of the replica partition function.

Secondly, our derivation of the monodromy method is based solely on CFT techniques and thus provides a non-trivial check of both the RT formula as well as previous calculations of \cite{Barrella:2013wja,Chen:2015kua} using the monodromy method on the gravity side.
Moreover, the monodromy method presented here is derived from first principles, further strengthening the heuristic arguments that were used to justify the monodromy method in \cite{Barrella:2013wja,Chen:2015kua}.
The derivation from the CFT side furthermore allows for a determination of the restrictions on the operator content and OPE coefficients that must be obeyed by the CFT in question in order for the vacuum block to give the dominant contribution and the RT results for the entanglement entropy to be valid.
Perhaps unsurprisingly, these restrictions are equivalent to those imposed in \cite{Hartman:2013mia} for the entanglement entropy at zero temperature.
That is, the entanglement entropy for any CFT with large central charge, sparse spectrum of light operators and at most exponentially growing OPE coefficients agrees with the RT formula not only at zero temperature as was already known from the results of \cite{Hartman:2013mia}, but also at finite temperature and finite size.
In addition, the same arguments that imply the universal form of the entanglement entropy in these CFTs also imply a universal form of the partition function at all temperatures, another feature of holographic CFT investigated also in \cite{Hartman:2014oaa}.

We close with a short outlook on possible future directions.
One interesting direction is an investigation in full generality of the issue under which conditions the twist correlator yields the same results as the replica partition function $Z_n$.
Previous work in the free fermion case \cite{Lokhande:2015zma,Mukhi:2017rex} indicates that for short intervals the twist correlator is equivalent to the replica partition function, a result which we can corroborate for large $c$ CFTs.
Does this hold in general?
Other directions can be found in further applications of the monodromy method.
Since conformal blocks are the basic building blocks of any correlation function, the derivation of the monodromy methods from first principles lays the groundwork for a number of applications.
In particular, the monodromy methods found in this paper may prove useful in deriving recursion relations in the dimensions of the exchanged operators for torus conformal blocks with better convergence properties than the known recursion relations in the central charge found in \cite{Cho:2017oxl}.
Moreover, the monodromy methods also pave the way to the study of other information theoretic quantities closely connected to the entanglement entropy such as entanglement negativity \cite{Calabrese:2012nk,Calabrese:2012ew,Calabrese:2014yza}, symmetry resolved entanglement \cite{Goldstein:2017bua,Bonsignori:2019naz,Murciano:2020lqq,Capizzi:2020,Zhao:2020qmn} or entwinement \cite{Balasubramanian:2014sra,Lin:2016fqk,Balasubramanian:2016xho,Balasubramanian:2018ajb,Erdmenger:2019lzr} for finite temperature states, providing opportunities for new insights into the physics of the dual AdS black holes.
We hope to be able to report on further results in this direction in the future.

\paragraph{Acknowledgements} I would like to thank Johanna Erdmenger and Christian Northe for useful discussions.
I acknowledge financial support by the Deutsche Forschungsgemeinschaft (DFG, German Research Foundation) under Germany's Excellence Strategy through Würzburg‐Dresden Cluster of Excellence on Complexity and Topology in Quantum Matter ‐ ct.qmat (EXC 2147, project‐id 390858490).

\appendix

\section{Conventions for elliptic functions}
\label{app:conventions}
This appendix contains an overview over conventions and useful identities for the Weierstraß elliptic functions used in the rest of the publication (see e.g.~\cite{NIST:DLMF} for more details about these functions).
All elliptic functions are defined for a lattice $\Lambda$ generated by the identifications $z \sim z + 1$ and $z \sim z + \tau$.
The Weierstraß elliptic functions $\wp(z),\zeta(z)$ and $\sigma(z)$ are defined by
\begin{equation}
  \begin{aligned}
    \wp(z) &= - \zeta'(z) = \frac{1}{z^2} + \sum_{(m,n) \neq (0,0)} \left(\frac{1}{(z+n+m\tau)^2} - \frac{1}{(n+m\tau)^2}\right),\\
    \zeta(z) &= \frac{\sigma'(z)}{\sigma(z)} = \frac{1}{z} + \sum_{(m,n) \neq (0,0)} \left(\frac{1}{(z+n+m\tau)} - \frac{1}{(n+m\tau)} + \frac{z}{(n+m\tau)^2}\right),\\
    \sigma(z) &= z \prod_{(m,n) \neq (0,0)} \left(1 - \frac{z}{n+m\tau}\right) \exp\left(\frac{z}{n+m\tau} + \frac{1}{2} \frac{z^2}{(n+m\tau)^2}\right).
  \end{aligned}
\end{equation}
$\wp(z)$ is a true elliptic (i.e.~doubly periodic) function while $\zeta(z)$ and $\sigma(z)$ are quasiperiodic:
\begin{equation}
  \begin{aligned}
    &\wp(z + 1) = \wp(z + \tau) = \wp(z)&\\
    &\zeta(z + 1) = \zeta(z) + 2\eta_1,  &\zeta(z + \tau) = \zeta(z) + 2\eta_3\\
    &\sigma(z + 1) = - \exp(2\eta_1(z + 1/2)) \sigma(z), &\sigma(z + \tau) = - \exp(2\eta_3(z + \tau/2)) \sigma(z),
  \end{aligned}
\end{equation}
where $\eta_1 = \zeta(1/2)$ and $\eta_3 = \zeta(\tau/2) = \tau\eta_1 - \pi i$.
Another useful definition of $\wp(z)$ and $\zeta(z)$ is given by
\begin{equation}
  \begin{aligned}
    \wp(z) + 2\eta_1 &= (2\pi i)^2 \sum_{m=-\infty}^\infty \frac{Q^m u}{(u - Q^m)^2}, \quad \wp(z)+\frac{2\eta_3}{\tau} = \left(\frac{2\pi i}{\tau}\right)^2 \sum_{m=-\infty}^\infty \frac{\tilde Q^m \tilde u}{(\tilde u - \tilde Q^m)^2},\\
    \zeta(z) - 2\eta_1 z &= i\pi \sum_{m=-\infty}^\infty \frac{Q^m + u}{Q^m - u}, \quad \zeta(z)-\frac{2\eta_3}{\tau} z = -\frac{i\pi}{\tau} \sum_{m=-\infty}^\infty \frac{\tilde Q^m + \tilde u}{\tilde Q^m - \tilde u},
  \end{aligned}
\end{equation}
where $u = e^{-2\pi i z}, Q = e^{2\pi i \tau}$ and $\tilde u = e^{2\pi i z/\tau}$, $\tilde Q = e^{-2\pi i/\tau}$.

\section{Recursion relations for torus conformal blocks}
\label{app:recursion-relations}
For completeness, this appendix shows the recursion formulas for the two-point conformal blocks on the torus following as a special case from the general method derived in \cite{Cho:2017oxl}.
For details of the derivation, see \cite{Cho:2017oxl} and also \cite{Hadasz:2009db} for the one-point torus block.
For the OPE block, the conformal block is given by the following recursion scheme
\begin{equation}
  \begin{aligned}
    \cF^\text{OPE}_{21,pq}(h_p,h_q,c) &= U^\text{OPE}(h_p,h_q,c)\\
    - \sum_{r \geq 2,s \geq 1}&\frac{\partial c_{rs}(h_q)}{\partial h_q} Q^{rs}\frac{A^{c_{rs}(h_q)}_{rs}\fusionpolynomial{c_{rs}(h_q)}{h_p}{h_q+rs}\fusionpolynomial{c_{rs}(h_q)}{h_p}{h_q}}{c-c_{rs}(h_q)}\cF^\text{OPE}_{21,pq}(h_q,h_q + rs,c_{rs}(h_q))\\
    - \sum_{r \geq 2,s \geq 1}&\frac{\partial c_{rs}(h_p)}{\partial h_p} y^{rs}\frac{A^{c_{rs}(h_p)}_{rs}\fusionpolynomial{c_{rs}(h_p)}{h_q}{h_q}\fusionpolynomial{c_{rs}(h_p)}{h_2}{h_1}}{c-c_{rs}(h_p)}\cF^\text{OPE}_{21,pq}(h_p + rs, h_q, c_{rs}(h_p))\\
  \end{aligned}
\end{equation}
where the fusion polynomials are given by
\begin{equation}
  \label{eq:fusionpolynomial}
  \fusionpolynomial{c}{h_1}{h_2} = \prod_{m = 1-r, m \in 1-r + 2\NN}^{r-1}\prod_{n = 1-s, n \in 1-s + 2\NN}^{s-1}\frac{\lambda_1+\lambda_2+mb+nb^{-1}}{2}\frac{\lambda_1-\lambda_2+mb+nb^{-1}}{2}
\end{equation}
with $\lambda_i = \sqrt{(b+b^{-1})^2-4h_i}$ while the prefactor is
\begin{equation}
  \label{eq:prefactor-fusionpolynomial}
  A^c_{rs} = \frac{1}{2}\prod_{\stackrel{(m,n)=(1-r,1-s)}{(m,n) \neq (0,0),(r,s)}}^{(r,s)}(mb+nb^{-1})^{-1}.
\end{equation}
$c_{rs}$ denotes the value of the central charge where degenerate representations of the Virasoro algebra appear,
\begin{equation}
  \begin{aligned}
    &c_{rs}(h) = 1 + 6(b_{rs}(h) + b_{rs}^{-1}(h))^2,\\
    &b_{rs}(h)^2 = \frac{rs - 1 + 2h + \sqrt{(r-s)^2+4(rs-1)h+4h^2}}{1-r^2}.
  \end{aligned}
\end{equation}
The $c$-regular part $U$ is given by
\begin{equation}
  U^\text{OPE}(h_p,h_q,c) = \left[\prod_{n=2}^\infty \frac{1}{1-Q^n}\right]\sum_{i,j \geq 0} Q^iy^j \frac{s_{ij}(h_q,h_p,h_q) (1-h_p-j)_{j}(h_p + h_1 - h_2)_{j}}{i!(2 h_q)_i j!(2 h_p)_j},
\end{equation}
where we define the rising and falling Pochhammer symbols by
\begin{equation}
  (a)_n = \prod_{k = 0}^{n-1} (a+k) \hspace{2cm} (a)^{(n)} = \prod_{k = 0}^{n-1} (a-k)
\end{equation}
and
\begin{equation}
  \begin{aligned}
    &s_{ij}(h_1,h_2,h_3) = \langle h_1| (L_{-1}^i)^\dagger \cO_{h_2}(1) L_{-1}^j |h_3 \rangle\\
    &= \left\{
      \begin{aligned}
        &\sum_{p = 0}^i \biggl(\arraycolsep=0pt\begin{array}{c}i \\ p \end{array}\biggr) (j)^{(p)} (2 h_3 + j-1)^{(p)}(h_1 + h_2 - h_3-(j-p))_{i-p}(h_3 + h_2 - h_1)_{j-p} ~,~ j \geq i\\
        &\sum_{p = 0}^j \biggl(\arraycolsep=0pt\begin{array}{c}j \\ p \end{array}\biggr) (i)^{(p)} (2 h_1 + i-1)^{(p)}(h_3 + h_2 - h_1-(i-p))_{j-p}(h_1 + h_2 - h_3)_{i-p} ~,~ i \geq j
      \end{aligned}\right.
  \end{aligned}
\end{equation}
The recursion formula for the projection block is given by \footnote{The authors of \cite{Cho:2017oxl} also claim to have found a different recursion formula in the conformal dimensions of the exchanged primaries $h_{p,q}$ for a class of conformal blocks termed the ``necklace blocks'' which include the projection block considered here as a special case. We cannot confirm this claim since the formula presented in \cite{Cho:2017oxl} agrees neither with the recursion relation in the central charge derived in \cite{Cho:2017oxl} nor with an explicit calculation in the first few orders of the series expansion in $q_1,q_2$ for the projection block.}
\begin{equation}
  \begin{aligned}
    \cF^\text{projection}_{2p,1q}(h_p,h_q,c) = &U^\text{projection}(h_p,h_q,c)\\
    - \sum_{r \geq 2,s \geq 1}&\frac{\partial c_{rs}(h_q)}{\partial h_q} q_1^{rs}\frac{A^{c_{rs}(h_q)}_{rs}\fusionpolynomial{c_{rs}(h_q)}{h_p}{h_1}\fusionpolynomial{c_{rs}(h_q)}{h_p}{h_2}}{c-c_{rs}(h_q)}\cF^\text{projection}_{2p,1q}(h_p,h_q + rs,c_{rs}(h_q))\\
    - \sum_{r \geq 2,s \geq 1}&\frac{\partial c_{rs}(h_p)}{\partial h_p} q_2^{rs}\frac{A^{c_{rs}(h_p)}_{rs}\fusionpolynomial{c_{rs}(h_p)}{h_q}{h_1}\fusionpolynomial{c_{rs}(h_p)}{h_q}{h_2}}{c-c_{rs}(h_p)}\cF^\text{projection}_{2p,1q}(h_p + rs,h_q,c_{rs}(h_p)),\\
  \end{aligned}
\end{equation}
where the $c$-regular part is given by
\begin{equation}
  U^\text{projection}(h_p,h_q,c) = \left[\prod_{n=2}^\infty \frac{1}{1-Q^n}\right]\sum_{i,j \geq 0} q_1^iq_2^j \frac{s_{ij}(h_q,h_2,h_q)s_{ji}(h_p,h_1,h_q)}{i!(2 h_q)_i j!(2 h_p)_j}.
\end{equation}
By explicit calculation, it is easy to check in the first few orders of the series expansion that for $h_1 = h_2$ and $h_{p,q} = \gamma c$, the limits $\lim_{\gamma \to 0}$ and  $\lim_{c \to \infty}$ of the OPE block commute.
We have checked this up to fourth order in $y,Q$.
A more convenient proof is possible with a recursion relation in the conformal weights of the exchanged operators $h_{p,q}$, as derived for the conformal block on the plane in \cite{Zamolodchikov1987}.
In fact, the singular parts proportional to $\sim 1/(h_{p,q} - h_{rs})$ of such a recursion relation are proportional to the singular parts $\sim 1/(c - c_{rs}(h_{p,q}))$ of the above recursion relations in $c$ \cite{Cho:2017oxl}.
Using these known singular parts, one can show that the limits $\gamma \to 0$ and $c \to \infty$ commute for the singular parts of this recursion relation to all orders, assuming the above conditions on $h_{1,2,p,q}$.
Together, these calculations provide some evidence that the semiclassical limit is well-defined for the vacuum block on the torus derived in \secref{sec:EE-thermal-states}.
Since the vacuum block on the torus in most limits is equivalent to a zero-point block on the replica surface $\cR_n$, this also provides evidence that the semiclassical limit for the zero-point vacuum block on $\cR_n$ is well-defined.

\bibliographystyle{JHEP}
\bibliography{bibliography}

\end{document}